\documentclass[reprint,amsmath,amssymb,aip,eqsecnum,jcp,longbibliography]{revtex4-1}
\usepackage[dvips]{graphicx}
\usepackage{color}
\usepackage{amsmath}
\usepackage{amssymb}
\usepackage{pstricks,pst-grad}
\usepackage{amsbsy}
\usepackage{epic}
\usepackage[normalem]{ulem}
\usepackage{hyperref}
\usepackage[hyphenbreaks]{breakurl}

\usepackage{dcolumn}% Align table columns on decimal point
\usepackage{bm}% bold math
\def\bal#1\eal{\begin{align}#1\end{align}}
\newcommand\beq{\begin{equation}}
\newcommand\eeq{\end{equation}}
\newcommand\beqa{\begin{eqnarray}}
\newcommand\eeqa{\end{eqnarray}}
\newcommand{\nn}{\nonumber\\}

\newcommand{\ex}{\text{ex}}
\newcommand{\cc}{\overline{c}}

\newcommand{\ed}{\end{document}}

\begin{document}

%\preprint{}


\title{Chemical potential of a test hard sphere of variable size in hard-sphere fluid mixtures}



\author{David M. Heyes}
\email{david.heyes@rhul.ac.uk}
\affiliation{Department of Physics, Royal Holloway, University of London, Egham, Surrey TW20 0EX, UK}
\author{Andr\'es Santos}
\email{andres@unex.es}
\homepage{http://www.unex.es/eweb/fisteor/andres/}
\affiliation{Departamento de F\'{\i}sica and Instituto de Computaci\'on Cient\'ifica Avanzada (ICCAEx), Universidad de Extremadura,
E-06071 Badajoz, Spain}

\date{\today}

\begin{abstract}
A detailed comparison between the Boubl\'ik--Mansoori--Carnahan--Starling--Leland (BMCSL)
equation of state of hard-sphere mixtures is made with Molecular Dynamics (MD) simulations of
the same compositions.
The Lab\'ik and Smith simulation technique [S. Lab\'ik and W. R. Smith, Mol. Simul. \textbf{12}, 23--31 (1994)]
was used to implement the Widom particle insertion method
to calculate the excess chemical potential, $\beta \mu_0^\text{ex}$, of
a test particle of variable diameter, $\sigma_0$, immersed in a hard-sphere fluid mixture with different compositions and values of the packing fraction, $\eta$.
Use is made of the fact that  the only polynomial representation
of $\beta \mu_0^\text{ex}$ which is consistent with the limits $\sigma_0\to 0$
and $\sigma_0\to\infty$
has to be of the cubic form, {i.e.},
$c_0(\eta)+\overline{c}_1(\eta)\sigma_0/M_{1}+\overline{c}_2(\eta)(\sigma_0/M_{1})^2+\overline{c}_3(\eta)(\sigma_0/M_{1})^3$, where $M_{1}$
is the first moment of the distribution. The first two coefficients, $c_0(\eta)$ and $\overline{c}_1(\eta)$, are known analytically, while
$\overline{c}_2(\eta)$ and $\overline{c}_3(\eta)$ were obtained by fitting the MD data to this expression.
This in turn provides a  method to determine the
excess free energy per particle, $\beta a^\text{ex}$, in terms of $\overline{c}_2$, $\overline{c}_3$, and the compressibility factor, $Z$.
Very good agreement between the
BMCSL formulas and the MD data is found for
$\beta \mu^\text{ex}_0$, $Z$, and $\beta a^\text{ex}$ for binary mixtures and continuous particle size distributions
with the top-hat analytic form. However, the BMCSL theory
typically  slightly underestimates the simulation values, especially for $Z$,
differences which  the Boubl\'ik--Carnahan--Starling--Kolafa formulas and an interpolation between two Percus--Yevick routes
capture well in different ranges of the system parameter space.

\end{abstract}

\date{\today}

%\keywords{}
\maketitle

%%%%%%%%%%%%%%%%%%%%%%%%%%%%%%%%%%%%%%%%%%%%%%%%%%%%%%%%%%%%%%%%%%%%%%%%%%%%%%%
% INTRODUCTION
%%%%%%%%%%%%%%%%%%%%%%%%%%%%%%%%%%%%%%%%%%%%%%%%%%%%%%%%%%%%%%%%%%%%%%%%%%%%%%%

\section{Introduction}

%Mixtures of hard particles
{Particulate mixtures are widely encountered in the real world, in the form of powders
and liquid mixtures. For many years, there has been an active interest in studying such ``granular'' mixtures in numerous fields, such as in chemistry, geology, pharmaceutical science, food technology,  and in various
aspects of chemical engineering processing and civil engineering. They are intrinsically
difficult to understand and control. The hard sphere (HS) particle has proved a useful reference fluid for single component liquids, and it makes logical sense to use the same type of model particle to
act as a starting point to represent  and understand the physical behavior of such multicomponent systems. Mixtures of HSs of different size distributions (discrete or continuous) can similarly be used to model, for instance,
%real multicomponent liquids of
%small molecules,
nanocolloidal liquids and granular materials.}
Mixtures are more problematic than single component liquids to deal with theoretically  as there are
more parameters to be accounted for in any theoretical treatment.
In the binary mixture HS case, these are the total
packing fraction, $\eta$,
the diameters of the two spheres, $\sigma_{1}$ and $\sigma_2$,
their mole fractions, $x_1$ and $x_2=1-x_1$, and for dynamical properties, their masses.
Therefore, the statistical mechanical theory of the equation of state (EoS) and derived
properties poses a much greater challenge than for the single component.  One of the most
widely used approximate analytic EoSs for mixtures is that of
Boubl\'ik--Mansoori--Carnahan--Starling--Leland (BMCSL).\cite{B70,MCSL71} This has been found to be an accurate representation of
the available simulation data.\cite{YCH96,BMLS96,L96,LW99,TT05,BS08,OB11} However, a comprehensive exploration
of its performance over the possible ranges
of this parameter space has yet
to be carried out. This is particularly for the chemical potential
of a test particle in regions of this parameter space which would better assess the overall
accuracy of the BMCSL EoS.\\

Here, a systematic exploration of the accuracy of the BMCSL EoS over a wide parameter range is
carried out using new computer simulation data of a binary mixture of HSs and a closely related so-called top-hat (TH)
continuous diameter distribution between  lower and higher values, $\sigma_{2}$ and $\sigma_{1}$, respectively.
The chemical potential of a test particle of varying diameter, $\sigma_{0}$, inserted in the  mixture
is calculated using the Widom particle insertion method.\cite{W63b}

The  method employed here was first used for HS systems by Monte Carlo (MC) simulations
for the single component case by Lab\'{i}k and Smith,\cite{LS94}
and later applied to a binary mixture of HSs by Baro\v{s}ov\'a {\it et al.}\cite{BMLS96}
The technique measures the probability of the successful insertion of
a test particle of arbitrary diameter $\sigma_0$. These measurements are extrapolated with a suitable polynomial
in powers of $\sigma_{0}$ to give the chemical potential of tracer particles larger than can be practically inserted in the simulations.
We first applied this method in Molecular Dynamics (MD) simulations to the single component fluid using a third-degree polynomial.\cite{HS16}
In the Appendix of Ref.~\onlinecite{HS16}, we proved that a polynomial consistent with the limits $\sigma_0\to 0$ and $\sigma_0\to\infty$ must necessarily have a cubic form, regardless of the number of components.
To be consistent
with this fact, we assume here  a  third-degree polynomial, even in the multicomponent case,  rather
than a fourth- or fifth-degree polynomial which was used in Ref.~\onlinecite{BMLS96}.
The highest two coefficients so obtained by fitting the simulation test particle chemical
potential data to the polynomial are used here to compute the free energy per particle of the mixture,
which is a novel outcome of this numerical study.
Recently, Baranau and Tallarek\cite{BT16} employed an alternative solution
which consisted of measuring the so-called pore-size distribution, fitting it to
a Gaussian, and then performing analytically an integral to determine the chemical potential. Other methods use a particle swap scheme\cite{H92} or well-tempered metadynamics.\cite{PGP16,VTP16}
We note in passing that
alternative analytic forms for the EoS  to the BMCSL equation have been proposed and
compared against  simulation data.\cite{SYH99,SYH01,SYH02,SYH05,HR06a,BS08,P15}.\\

The remainder of this work is organized as follows. Expressions are reviewed in Sec.~\ref{sec2} and the use of a cubic polynomial as a trial function for $\beta\mu_0^\ex$ is justified. The computer simulation method is briefly described in Sec.\ \ref{sec2CS}.
The results are presented and compared with theoretical predictions in Sec.~\ref{sec3}, and
some conclusions are given in Sec.~\ref{sec4}.

\section{Theory}
\label{sec2}

\subsection{Equation of state of multicomponent hard-sphere fluids}

Consider a three-dimensional fluid mixture of additive HSs with
an arbitrary number of components, in which for each species $j$, there are $N_j$  spheres of diameter $\sigma_j$.
The total number of particles is $N=\sum_j N_j$ and the $n$th moment of the size distribution is
\beq
M_n=\sum_{j}x_j\sigma_j^n,
\label{2.2X}
\eeq
where $x_j=N_j/N$ is the mole fraction of species $j$ (with $\sum_j x_j=1$).
The total packing or volume fraction of the HS mixture is exactly defined as
\beq
\eta=\frac{\pi}{6}\frac{N}{V}M_3,
\label{etaX}
\eeq
where $V$ is the volume of the system.

The compressibility factor of the mixture is denoted by $Z(\eta,\{x_j\})\equiv \beta p V/N$,
where $p$ is the pressure, $\beta=1/k_{B}T$,  $k_{B}$ is Boltzmann's constant, and $T$ is the absolute temperature.
The exact form  of $Z$ as a function of these parameters is not known,  and several approximations have been proposed.\cite{MGPC08,BS08}

The exact solution\cite{LZ71,PS75,B75} of the Percus--Yevick (PY) integral equation\cite{PY58} leads to explicit expressions for $Z$ by different thermodynamic routes. Specifically, the virial (PY-v), compressibility (PY-c), and chemical-potential (PY-$\mu$) routes in the PY approximation
have a common structure,\cite{S16,LZ71,PS75,B75,S12b,SR13}
\bal
Z(\eta,\{x_j\})=&Z_0(\eta)+Z_1(\eta)\frac{M_1M_2}{M_3}+Z_2(\eta)\frac{M_2^3}{M_3^2}\nn
=&Z_0(\eta)+Z_1(\eta)\lambda+Z_2(\eta)\frac{\lambda^3}{\gamma},
\label{2.1}
\eal
where
\beq
Z_0(\eta)=\frac{1}{1-\eta},\quad
Z_1(\eta)=\frac{3\eta}{(1-\eta)^2},
\label{2.3}
\eeq
\beq
\lambda(\{x_j\})\equiv \frac{M_1M_2}{M_3},\quad \gamma(\{x_j\})\equiv\frac{M_1^3}{M_3}.
\eeq
Of course, $\lambda=\gamma=1$ is the single component case.
In general, as proved in Ref.\ \onlinecite{OL12},  one has $\lambda^2\leq\gamma\leq \lambda\leq 1$, regardless of the number of components.   The functions \eqref{2.3} are the same in all the PY EoSs, while
the function $Z_2(\eta)$ depends on the route and is displayed in the second column of Table \ref{table1}.
It must be noted that the PY-c EoS is equivalent to the Scaled Particle Theory (SPT) approximation,\cite{RFL59,LHP65,MR75,R88,HC04b} and this is why it is labeled as ``PY-c (SPT)'' in Table \ref{table1}.

%\begin{squeezetable}
  \begin{table*}
   \caption{Expressions for $Z_2(\eta)$, $a_2(\eta)$, $c_2(\eta)$,  and $c_3(\eta)$, according to several approximations.}\label{table1}
\begin{ruledtabular}
\begin{tabular}{lcccc}
Approx.&$Z_2(\eta)$&$a_2(\eta)$&$c_2(\eta)$&$c_3(\eta)$\\
\hline
PY-v&$\displaystyle{\frac{3\eta^2}{(1-\eta)^2}}$&
$\displaystyle{3\ln(1-\eta)+\frac{3\eta}{1-\eta}}$&$\displaystyle{9\ln(1-\eta)+12\frac{\eta}{1-\eta}}$&
$\displaystyle{-6\ln(1-\eta)-\eta\frac{5-11\eta}{(1-\eta)^2}}$\\
PY-c (SPT)&$\displaystyle{\frac{3\eta^2}{(1-\eta)^3}}$&
$\displaystyle{\frac{3\eta^2}{2(1-\eta)^2}}$&$\displaystyle{3\eta\frac{2+\eta}{2(1-\eta)^2}}$&
$\displaystyle{\eta\frac{1+\eta+\eta^2}{(1-\eta)^3}}$\\
PY-$\mu$&$\displaystyle{-\frac{9\ln(1-\eta)}{\eta}-9\frac{1-\frac{3}{2}\eta}{(1-\eta)^2}}$&
$\displaystyle{\frac{9\ln(1-\eta)}{\eta}+9\frac{1-\frac{1}{2}\eta}{1-\eta}}$&$\displaystyle{27\frac{\ln(1-\eta)}{\eta}+3\frac{18-7\eta}{2(1-\eta)}}$&
$\displaystyle{-27\frac{\ln(1-\eta)}{\eta}-\frac{54-83\eta+14\eta^2}{2(1-\eta)^2}}$\\
BMCSL&$\displaystyle{\frac{\eta^2(3-\eta)}{(1-\eta)^3}}$&$\displaystyle{\ln(1-\eta)+\frac{\eta}{(1-\eta)^2}}$&$\displaystyle{3\ln(1-\eta)+3\eta\frac{2-\eta}{(1-\eta)^2}}$&$\displaystyle{-2\ln(1-\eta)-\eta\frac{1-6\eta+3\eta^2}{(1-\eta)^3}}$\\
BCSK&$\displaystyle{\frac{\eta^2[3-\frac{2}{3}\eta(1+\eta)]}{(1-\eta)^3}}$&
\hspace{-1cm}$\displaystyle{\frac{8}{3}\ln(1-\eta)}$&$\displaystyle{8\ln(1-\eta)+\eta\frac{22-21\eta+4\eta^2}{2(1-\eta)^2}}$&
$\hspace{-1.6cm}\displaystyle{-\frac{16}{3}\ln(1-\eta)}$\\
 & & $\displaystyle{\qquad+\eta\frac{16-15\eta+4\eta^2}{6(1-\eta)^2}}$&&$\displaystyle{\qquad-\eta\frac{13-43\eta+27\eta^2-2\eta^3}{3(1-\eta)^3}}$\\
\end{tabular}
 \end{ruledtabular}
 \end{table*}
%\end{squeezetable}

Equation \eqref{2.1} provides a method to extend any single component HS compressibility factor, $Z_s(\eta)$, to multicomponent fluids simply by choosing
\beq
\label{Zs}
Z_2(\eta)=Z_s(\eta)-\frac{1+2\eta}{(1-\eta)^2},
\eeq
although other alternative methods are possible.\cite{BS08,HYS08,SYH11,S12c,SYHOO14}
In particular, if $Z_s(\eta)$ is chosen
to be
the Carnahan--Starling\cite{CS69} or the Carnahan--Starling--Kolafa\cite{BN86} EoSs, application of Eqs.\ \eqref{2.1} and \eqref{Zs} yields the BMCSL or Boubl\'ik--Carnahan--Starling--Kolafa  (BCSK)\cite{B86} extensions, respectively. The corresponding expressions for $Z_2(\eta)$ are also presented in Table \ref{table1}.

As is well known, the BMCSL compressibility factor is
an interpolation between the PY-v and PY-c prescriptions, namely
\beq
Z^{\text{BMCSL}}(\eta)=\frac{1}{3}Z^{\text{PY-v}}(\eta)+\frac{2}{3}Z^{\text{PY-c}}(\eta).
\eeq
Given that the PY-$\mu$ route is slightly more accurate than the PY-v one,\cite{S12b,SR13} an alternative interpolation formula is
\beq
\label{PYmuc}
Z^{\text{PY-$\mu$c}}(\eta)=\alpha Z^{\text{PY-$\mu$}}(\eta)+(1-\alpha)Z^{\text{PY-c}}(\eta),
\eeq
with $\alpha\simeq 0.37$.

\subsection{Free energy of multicomponent hard-sphere fluids}
The thermodynamic relation between the excess free energy per particle, $a^\ex(\eta,\{x_j\})$, and the compressibility factor, $Z(\eta,\{x_j\})$, is
\beq
\beta a^\ex(\eta,\{x_j\})= \int_0^1 dt\frac{Z(\eta t,\{x_j\})-1}{t}.
\label{2.4}
\eeq
Therefore, if $Z$ has the form in Eq.~\eqref{2.1}, then
\beq
\beta a^\ex(\eta,\{x_j\})=c_0(\eta)+c_1(\eta)\lambda+a_2(\eta)\frac{\lambda^3}{\gamma},
\label{aex}
\eeq
where
\begin{subequations}
\label{2.6&2.4b}
\beq
c_0(\eta)=-\ln(1-\eta),\quad c_1(\eta)=\frac{3\eta}{1-\eta},
\label{2.6.0}
\eeq
\beq
a_2(\eta)= \int_0^1 dt\frac{Z_2(\eta t)}{t}.
\label{2.4b}
\eeq
\end{subequations}
The expressions for $a_2(\eta)$ corresponding to several approximations are also presented in Table \ref{table1}.

\subsection{Chemical potential of a test sphere}

Now, we want to obtain from Eq.\ \eqref{aex} the theoretical excess chemical potential of a test particle (of diameter $\sigma_0$) immersed in a HS mixture. To that end, note first that  the excess chemical potential of a generic species $i$  is thermodynamically defined as
\beq
\beta\mu_i^\ex=\left(\frac{\partial N \beta a^\ex}{\partial N_i}\right)_{V,N_{j\neq i}}.
\label{2.7}
\eeq
Next, $\beta\mu_0^\ex$ is obtained from $\beta\mu_i^\ex$ by the replacement $\sigma_i\to\sigma_0$ while keeping the composition of the mixture fixed. From Eq.\ \eqref{aex} one finally obtains\cite{HS16}
\beq
\beta \mu_0^\ex(\sigma_0)=c_0+\cc_1\frac{\sigma_0}{M_1}+\cc_2\left(\frac{\sigma_0}{M_1}\right)^2
+\cc_3\left(\frac{\sigma_0}{M_1}\right)^3,
\label{2.11}
\eeq
where
\begin{subequations}
\label{cc1-cc3}
\beq
\cc_1=\lambda\frac{3\eta}{1-\eta},
\label{cc1}
\eeq
\beq
\cc_2=\lambda^2 c_2+(\gamma-\lambda^2)\frac{3\eta}{1-\eta},
\label{cc2}
\eeq
\beq
\cc_3=\lambda^3 c_3+(\gamma-\lambda^3)\frac{\eta}{1-\eta}+\lambda(\gamma-\lambda^2)\frac{3\eta^2}{(1-\eta)^2}.
\label{cc3}
\eeq
\end{subequations}
In Eqs.\ \eqref{cc2} and \eqref{cc3},
\begin{subequations}
\label{2.14.1&2.14}
\beq
\label{2.14.1}
c_2(\eta)=\frac{3\eta}{1-\eta}+3a_2(\eta),
\eeq
\beq
c_3(\eta)=\frac{3\eta}{(1-\eta)^2}+Z_2(\eta)-\frac{2}{3}c_2(\eta).
\label{2.14}
\eeq
\end{subequations}
The expressions for $c_2(\eta)$ and $c_3(\eta)$ predicted by several approximations are given in Table \ref{table1}.
{}From Eqs.\ \eqref{2.1}, \eqref{2.3}, \eqref{cc1-cc3}, and \eqref{2.14.1&2.14} one can obtain the relationship
\beq
Z(\eta,\{x_j\})=1+\frac{1}{3}\cc_1+\frac{2}{3}\frac{\lambda\cc_2}{\gamma}+\frac{\cc_3}{\gamma}.
\label{LSeq}
\eeq

As mentioned above,
the structure of Eq.\ \eqref{2.1}, and hence of Eqs.\ \eqref{aex}, and \eqref{2.11}, is common to several approximate EoSs, such as PY-v, PY-c, PY-$\mu$, BMCSL, BCSK, and, obviously, PY-$\mu$c.
They all share the coefficients $Z_0$, $Z_1$, $c_0$, and $c_1$ [see Eqs.\ \eqref{2.3} and \eqref{2.6.0}], but differ in $Z_2$, $a_2$, $c_2$, and $c_3$. Only $Z_2$ is an independent quantity, since $a_2$, $c_2$, and $c_3$ are given by Eqs.\ \eqref{2.4b}, \eqref{2.14.1}, and \eqref{2.14}, respectively.

{An analysis, in the case of binary mixtures, of the extremal properties of the combinations of $\lambda$ and $\gamma$ appearing in Eqs.\ \eqref{2.1} and \eqref{cc1-cc3} is presented in the Appendix.}

\subsection{Consistency conditions in the limits $\sigma_0\to 0$ and $\sigma_0\to\infty$}
As proved in the Appendix of Ref.\ \onlinecite{HS16}, the exact form of $\beta\mu_0^\ex$ in the limit $\sigma_0/M_1\ll 1$ is
\beq
\beta \mu_0^\ex(\sigma_0)=c_0+\cc_1\frac{\sigma_0}{M_1}+\mathcal{O}\left(({\sigma_0}/{M_1})^2\right).
\label{2.11asympt}
\eeq
Therefore, the cubic approximation \eqref{2.11} is
consistent with the asymptotic behavior \eqref{2.11asympt}. Moreover,
it is noteworthy that if $\beta \mu_0^\ex$ is represented
by a polynomial in the diameter $\sigma_0$, the polynomial must \emph{necessarily} be of third degree.
This is a consequence of the physical requirement that, in the limit of an infinitely large impurity,
the condition\cite{RFHL60,RELK02,S12c}
\beq
\eta Z=M_3\lim_{\sigma_0\to\infty}\frac{\beta\mu_0^\ex(\sigma_0)}{\sigma_0^3}
\label{2b.4}
\eeq
must be obeyed.
Therefore, since $\lim_{\sigma_0\to\infty}{\beta\mu_0^\ex(\sigma_0)}/{\sigma_0^3}$ can be neither zero nor infinity, the only polynomial approximations consistent with that property are those of third degree.

In the case of the approximations of the form \eqref{2.1}, Eq.~\eqref{2b.4} implies
\beq
Z=\frac{1}{\eta}\frac{\cc_3}{\gamma}.
\label{2c.5}
\eeq
However, the PY-v, PY-$\mu$, BMCSL, and BCSK EoSs are not fully consistent with Eq.~\eqref{2c.5}.
This means that those approximations \emph{qualitatively} agree with the physical requirement \eqref{2b.4} since $\lim_{\sigma_0\to\infty}{\beta\mu_0^\ex(\sigma_0)}/{\sigma_0^3}=\text{finite}$ but yield different results for the left- and right-hand sides. The difference between $\eta Z$, as given by Eqs.\ \eqref{2.1} or \eqref{LSeq}, and $\cc_3/\gamma$ can be
seen to be
\beq
\eta Z-\frac{\cc_3}{\gamma}=\frac{\lambda^3}{\gamma}\left[2a_2-\eta(1-\eta)a_2'\right],
\eeq
where $a_2'(\eta)=d a_2(\eta)/d\eta$.
Thus, a perfect agreement between Eqs.\ \eqref{LSeq} and \eqref{2c.5}  is only possible if $a_2'=2a_2/\eta(1-\eta)$, whose general solution is $a_2=K\eta^2/(1-\eta)^2$, where $K$ is a constant. The associated one-component third virial coefficient is $b_3=7+2K$, so that $b_3=10$ implies $K=\frac{3}{2}$ and thus one recovers the PY-c (SPT) approximation.

\subsection{Equivalence between different mixtures}
According to Eqs.\ \eqref{2.1}, \eqref{aex}, and \eqref{cc1-cc3}, two mixtures sharing the same values of $\eta$, $\lambda$, and $\gamma$ would have common values of $Z$, $\beta a^\ex$, and $\cc_n$.
Having the same values of $\lambda$ and $\gamma$ implies having the same values of the
reduced moments $M_2/M_1^2$ and $M_3/M_1^3$. As a consequence, given any HS mixture
(with an arbitrary number of components and arbitrary diameters), it
will
always
be possible to find an ``equivalent'' binary mixture with the same equilibrium
properties. The mole fraction, $x_1$, of the bigger spheres and the size
ratio, $q\equiv \sigma_2/\sigma_1\leq 1$, for the binary mixture equivalent
of a multicomponent system characterized by given values of $\lambda$ and $\gamma$
are obtained using the formulas
\begin{subequations}
\label{x1&q_equiv}
\beq
\label{x1_equiv}
x_1=\frac{1}{2}-\frac{1+2 \gamma -3 \lambda  }{2 \sqrt{1+4 \gamma -\lambda(6+3   \lambda  -4 \lambda ^2/\gamma)}},
\eeq
\beq
\label{q_equiv}
q=\frac{ 1-\lambda-\sqrt{1+4 \gamma -\lambda(6+3   \lambda  -4 \lambda ^2/\gamma)} }{2 \left[\lambda(1+\lambda -\lambda ^2/\gamma)-\gamma  \right]/ (1-\lambda )}-1.
\eeq
\end{subequations}

{The mapping property discussed above applies as well to a continuous size distribution characterized by a given distribution function $x(\sigma)$, in which case the moments are
\begin{equation}
  M_n=\int_0^\infty d\sigma\, x(\sigma) \sigma^n.
\end{equation}
In particular, let us consider  a TH continuous size distribution
of HSs with
diameters between $\sigma_{\min}$ and $\sigma_{\max}$:
\begin{equation}
  x_{\text{TH}}(\sigma)=\frac{1}{\sigma_{\max}-\sigma_{\min}}\begin{cases}
    1&\text{if }\sigma_{\min}<\sigma<\sigma_{\max},\\
    0&\text{otherwise}.
  \end{cases}
\end{equation}
 The associated values of $\lambda$ and $\gamma$ are
\beq
\label{lambdaTH}
\lambda_{\text{TH}}=\frac{2}{3}\frac{1+q_{\text{TH}}+q_{\text{TH}}^2}{1+q_{\text{TH}}^2},\quad \gamma_{\text{TH}}=\frac{1}{2}\frac{\left(1+q_{\text{TH}}\right)^2}{1+q_{\text{TH}}^2},
\eeq
where $q_{\text{TH}}=\sigma_{\min}/\sigma_{\max}\leq 1$ is the size ratio.}
Insertion of Eq.\ \eqref{lambdaTH} into Eqs.\ \eqref{x1&q_equiv}  yields $x_1=\frac{1}{2}$ and
\beq
\label{qTH}
q=\frac{q_{\text{TH}}+2-\sqrt{3}}{1+(2-\sqrt{3})q_{\text{TH}}}.
\eeq
For instance, if $q_{\text{TH}}=0$, the equivalent binary mixture must have a size ratio $q=2-\sqrt{3}\simeq 0.268$. Inversion of Eq.\ \eqref{qTH} gives
\beq
\label{qTH_from_q}
q_{\text{TH}}=\frac{q-(2-\sqrt{3})}{1-(2-\sqrt{3})q}.
\eeq

\section{Computer Simulations}
\label{sec2CS}

The application of MD and MC computer simulation to HS mixtures goes back to
the 1960s.\cite{A64,R65}
The MD simulations carried out in this study employed a generalization of the
methodology used in our previous study of single component HSs,\cite{HS16} using the equations
relevant to binary, and hence to multicomponent HS mixtures in general, given by
Bannerman and Lue.\cite{BL09}
{A test point particle was randomly inserted in the system.
If it did not fall within an existing sphere, the distance, $r_{n}$, from the point to the center of the nearest sphere, of type $k$ and diameter $\sigma_{k}$,
was computed. All the values of $\sigma_{0}=0$ to $\sigma_0=2r_{n}-\sigma_{k}$ would be allowed insertions which were correspondingly logged.}
For HSs the Widom method reduces
to a simple bookkeeping procedure as the Boltzmann factor is either  $1$, if the
test sphere (and hence other test particles with diameter less than $\sigma_{0}$) does not overlap with any of the $N$ particles, or is equal to $0$, if there is an overlap with any of the other spheres.

The simulations were carried out employing $N=2048$ particles for typically $2 \times 10^{5}$
collisions per particle. Some additional simulations with $N=4000$ particles were performed for several of
the  mixtures to assess the influence of finite-size effects. The pressure and hence $Z$ were obtained by the virial theorem written in terms of the collision parameter.
The thermodynamics of the systems is independent of the masses of the spheres, but to facilitate equilibration of
the smaller particles,  the mass was set to  $m_{i}\propto\sigma_{i}$ for the binary mixtures
and  $m(\sigma)\propto\sqrt{\sigma}$ for the TH distributions. The simulations
were carried out in the fluid phase diagrams, as
these regions are already known for binary\cite{BH91,DRE99b} and
polydisperse\cite{KB99}  mixtures.

\begin{figure}[tbp]
\includegraphics[width=\columnwidth]{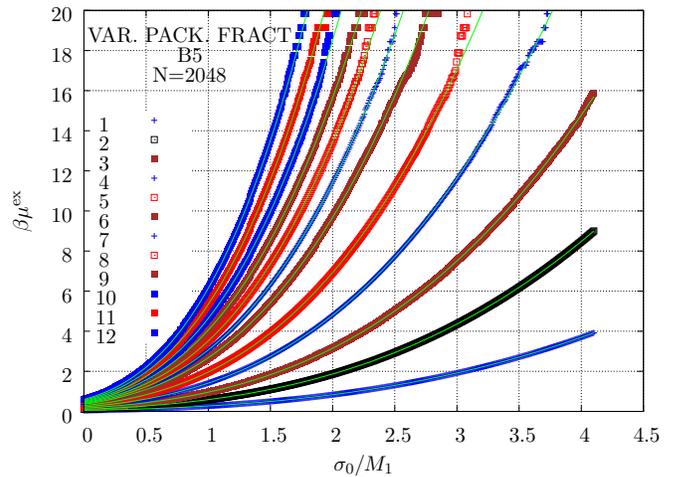}
\caption{Plot of the excess chemical potential of a test particle,
$\beta \mu_{0}^{\ex}(\sigma_0)$, for the B5 type of system (see Table \ref{table3} below). From right to left, the curves are for $\eta=0.05, 0.1, 0.15, 0.2, 0.25, 0.3, 0.325, 0.35, 0.375,  0.4, 0.425, $ and $0.45$, respectively, using $N=2048$ particles in the MD simulation.} \label{fig:fit}
\end{figure}

For each mixture,  the excess chemical potential, $\beta\mu_0^\ex$,
was computed by the Widom method as a function of the dimensionless impurity diameter $\sigma_0/M_1$.
$\mathcal{N}$ attempted particle insertions were made randomly in the system at the same time after every $\mathcal{N}$ collisions.
These data were fitted to a cubic polynomial of the form \eqref{2.11}
using the known \emph{exact}  values  for $c_0$ and  $\cc_1$ from
the expressions in Eqs.\ \eqref{2.6.0} and \eqref{cc1}, respectively.
In fact, $c_0$ is \emph{universal} in the sense that it is independent of the details of the mixture, apart from the
total packing fraction.
The parameters, $\cc_2$ and $\cc_3$ were obtained by fitting the  simulation
values of $\beta\mu_0^\ex$ to the cubic polynomial in $\sigma_0/M_1$.
The fitting procedure was applied to values $\beta\mu_0^\ex\leq 16.0$, which corresponds to insertion probabilities larger than $\sim 10^{-7}$. An example of the quality of fit is shown in Fig. \ref{fig:fit}.

From the values of $Z$, $\cc_2$, and $\cc_3$ (and hence $\beta\mu_i^\ex$)  obtained from the simulations, the excess free energy per particle
was obtained from the Euler equation of thermodynamics, i.e.,
\bal
\label{3.1}
\beta a^\ex=&\sum_i x_i\beta\mu_i^\ex-Z+1\nn
=&-\ln(1-\eta)+\lambda\frac{3\eta}{1-\eta}+ \frac{\lambda \cc_2}{\gamma}+\frac{\cc_3}{\gamma}-Z+1,
\eal
where in the last step use has been made of Eq.\ \eqref{2.11}.

\begin{table}[tbp]
\caption{The HS mixture systems analyzed and their tags.}\label{table3}
\begin{ruledtabular}
\begin{tabular}{cccc}
Set & $\eta$&$x_1$&$\sigma_2/\sigma_1$\\
\hline
B1&$0.3$&$0.5$&Variable\\
B2&$0.3$&Variable&$0.5$\\
B3&$0.3$&Variable&$0.3$\\
B4&Variable&$0.5$&$0.5$\\
B5&Variable&$0.5$&$2-\sqrt{3}$\\
B6&$(x_1+0.1)/2$&Variable&$x_1$\\
\hline
Set & $\eta$&\multicolumn{2}{c}{$\sigma_{\min}/\sigma_{\max}$}\\
\hline
TH1&$0.3$&\multicolumn{2}{c}{Variable}\\
TH2&Variable&\multicolumn{2}{c}{$0$}\\
\end{tabular}
\end{ruledtabular}
\end{table}

\begin{figure}[tbp]
\includegraphics[width=7cm]{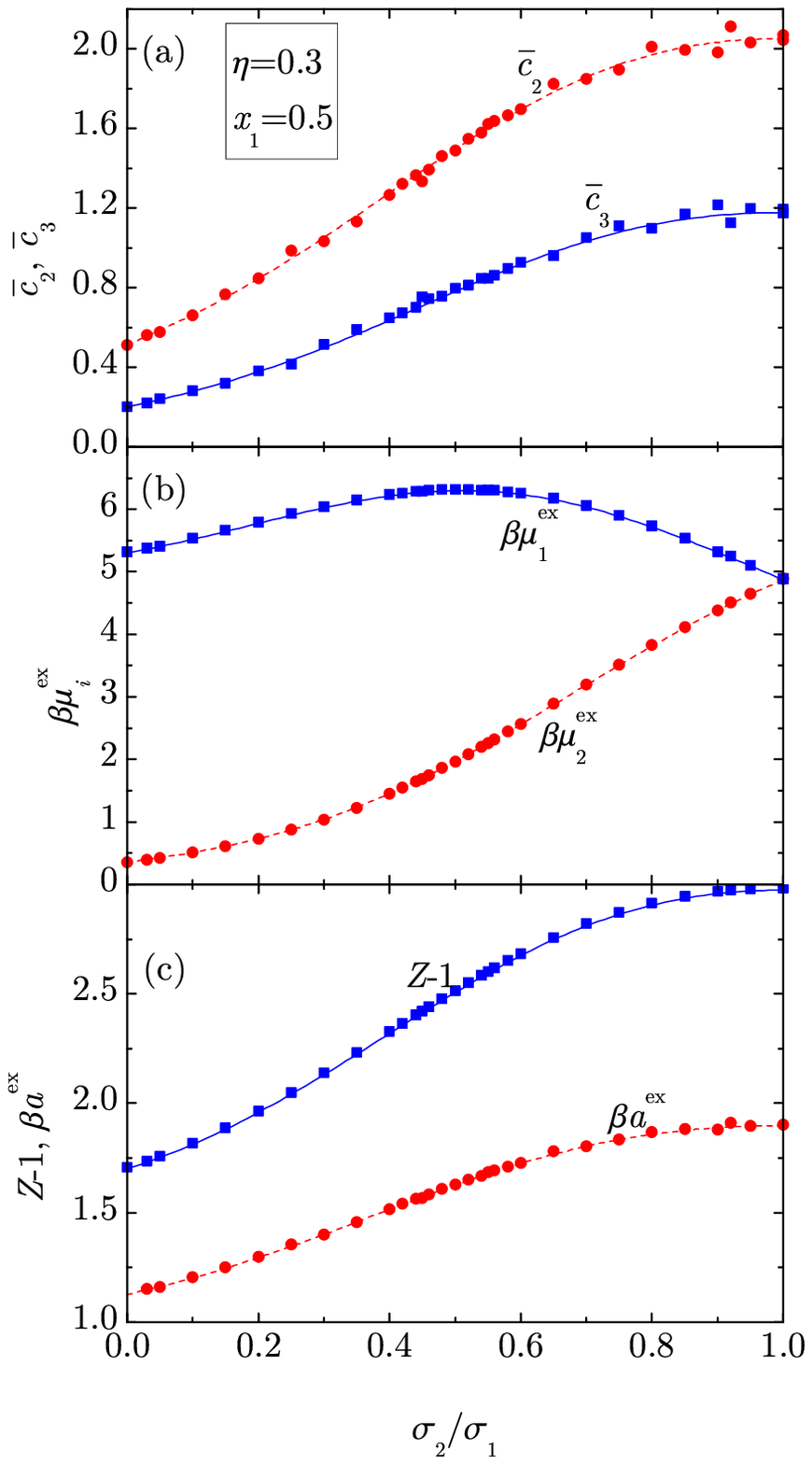}
\caption{Plot of (a) $\cc_2$ and $\cc_3$, (b) $\beta\mu_1^\ex$ and $\beta\mu_2^\ex$, and (c) $Z-1$ and $\beta a^\ex$ versus $\sigma_2/\sigma_1$ for system B1. The symbols are our MD data ($N=2048$) and the lines are the BMCSL predictions.\label{fig:B1}}
\end{figure}

\begin{figure}[tbp]
\includegraphics[width=7cm]{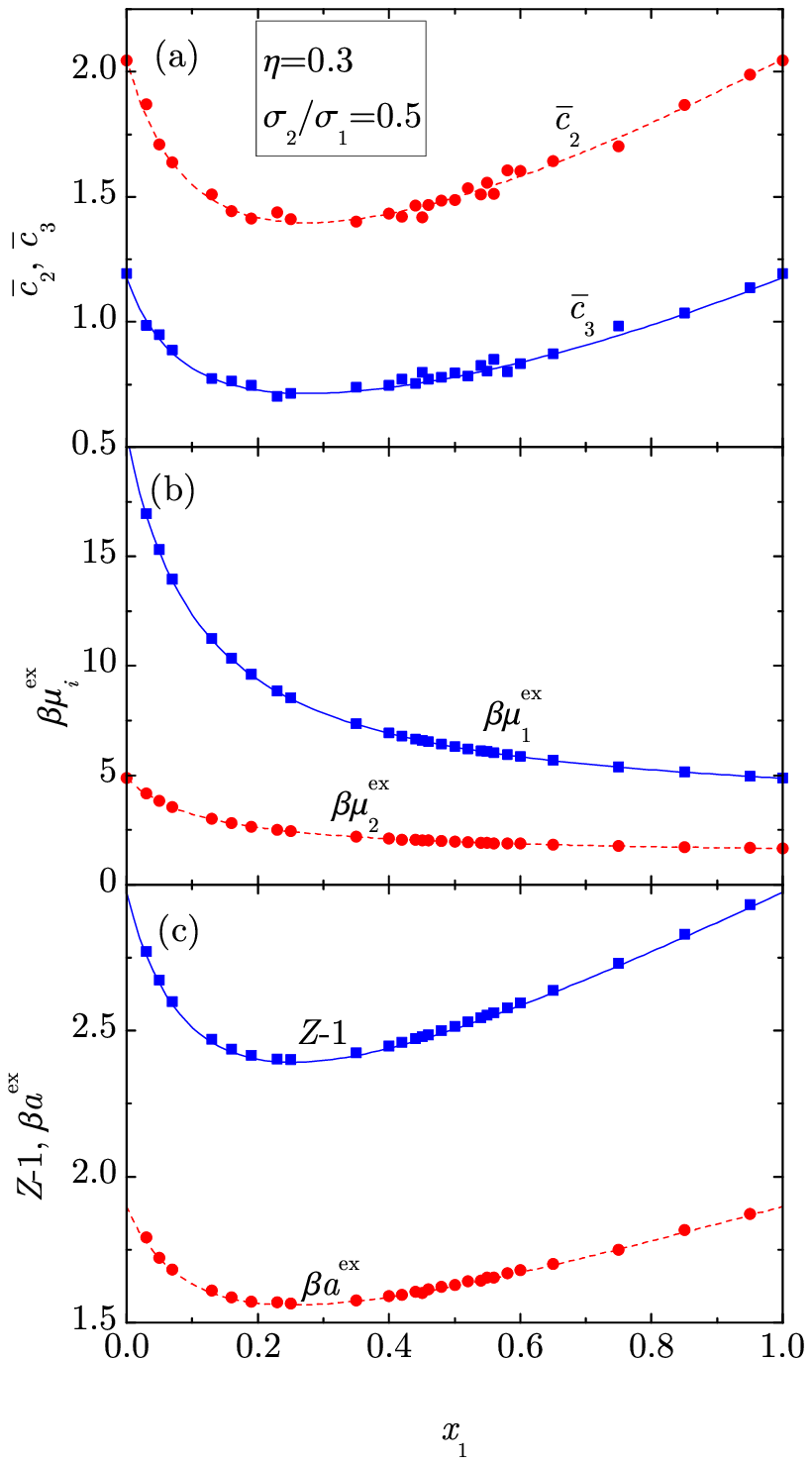}
\caption{Plot of (a) $\cc_2$ and $\cc_3$, (b) $\beta\mu_1^\ex$ and $\beta\mu_2^\ex$, and (c) $Z-1$ and $\beta a^\ex$ versus $x_1$ for system B2. The symbols are our MD data ($N=2048$) and the lines are the BMCSL predictions.\label{fig:B2}}
\end{figure}

\begin{figure}[tbp]
\includegraphics[width=7cm]{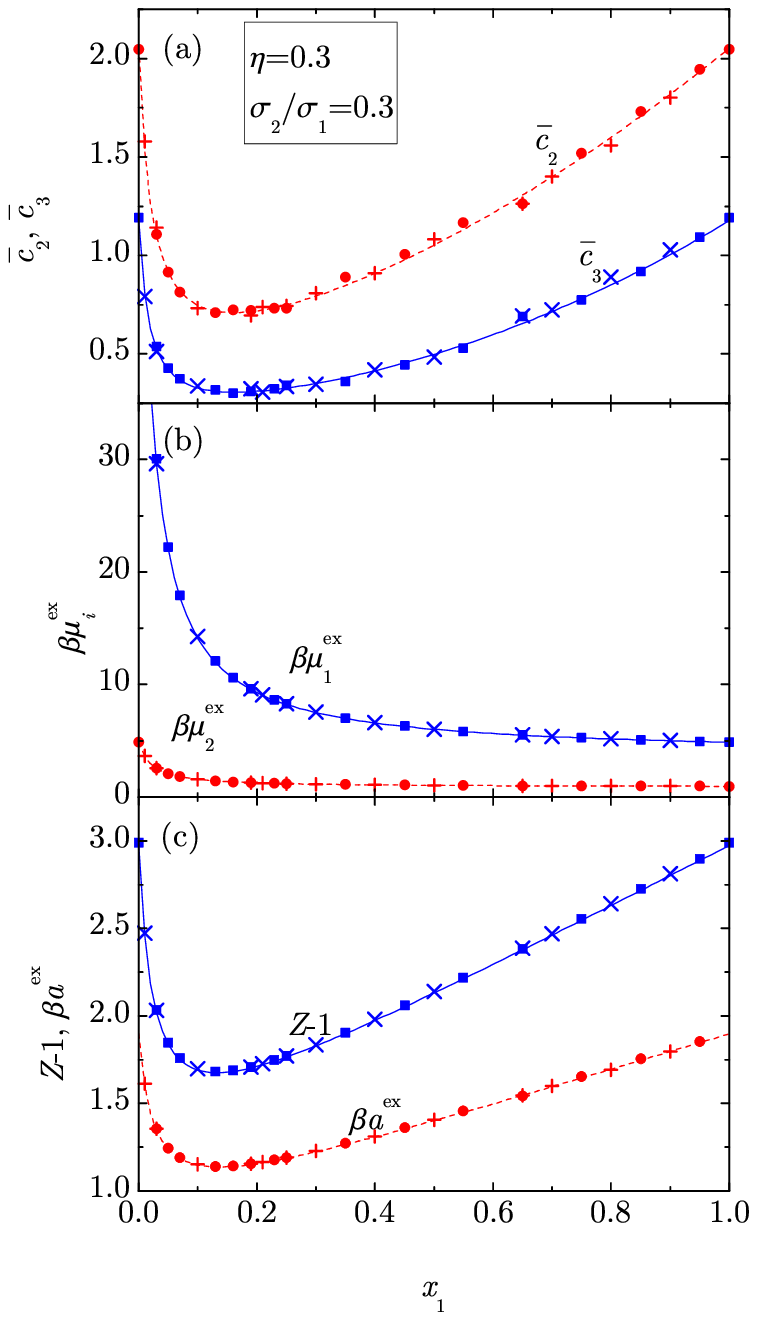}
\caption{Plot of (a) $\cc_2$ and $\cc_3$, (b) $\beta\mu_1^\ex$ and $\beta\mu_2^\ex$, and (c) $Z-1$ and $\beta a^\ex$ versus $x_1$ for system B3. The symbols are our MD data (filled symbols: $N=2048$, crosses: $N=4000$) and the lines are the BMCSL predictions.\label{fig:B3}}
\end{figure}

\begin{figure}[tbp]
\includegraphics[width=6.5cm]{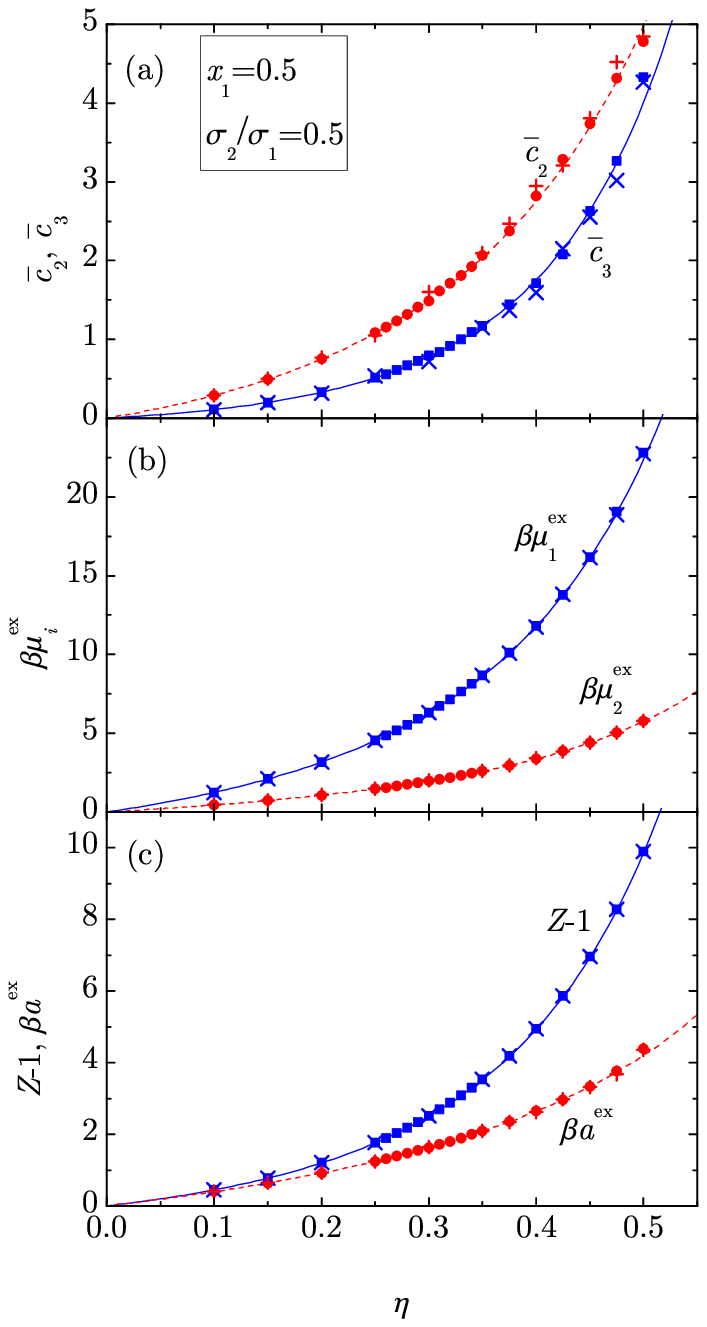}
\caption{Plot of (a) $\cc_2$ and $\cc_3$, (b) $\beta\mu_1^\ex$ and $\beta\mu_2^\ex$, and (c) $Z-1$ and $\beta a^\ex$ versus $\eta$ for system B4. The symbols are our MD data (filled symbols: $N=2048$, crosses: $N=4000$) and the lines are the BMCSL predictions.\label{fig:B4}}
\end{figure}

\begin{figure}[tbp]
\includegraphics[width=7cm]{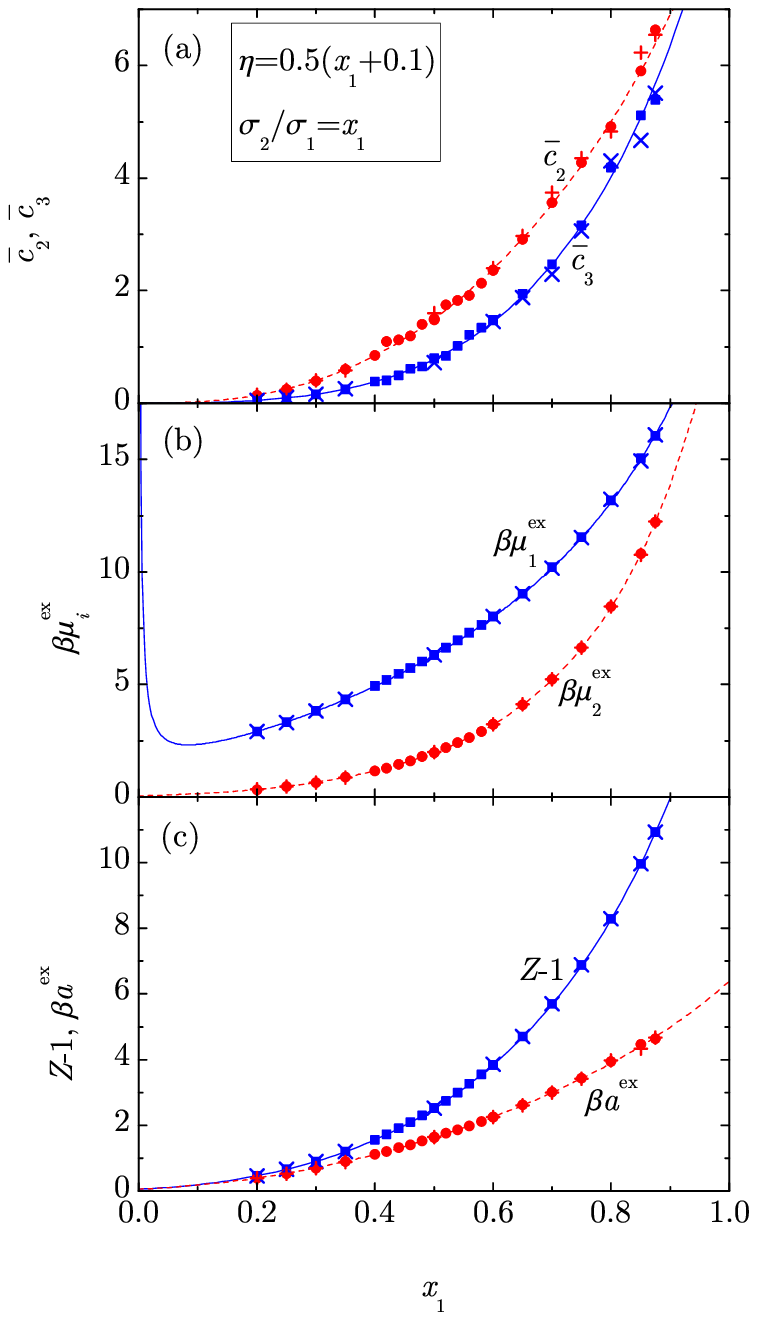}
\caption{Plot of (a) $\cc_2$ and $\cc_3$, (b) $\beta\mu_1^\ex$ and $\beta\mu_2^\ex$, and (c) $Z-1$ and $\beta a^\ex$ versus $x_1$ for system B6. The symbols are our MD data (filled symbols: $N=2048$, crosses: $N=4000$) and the lines are the BMCSL predictions.\label{fig:B6}}
\end{figure}

\begin{figure}[tbp]
\includegraphics[width=6.8cm]{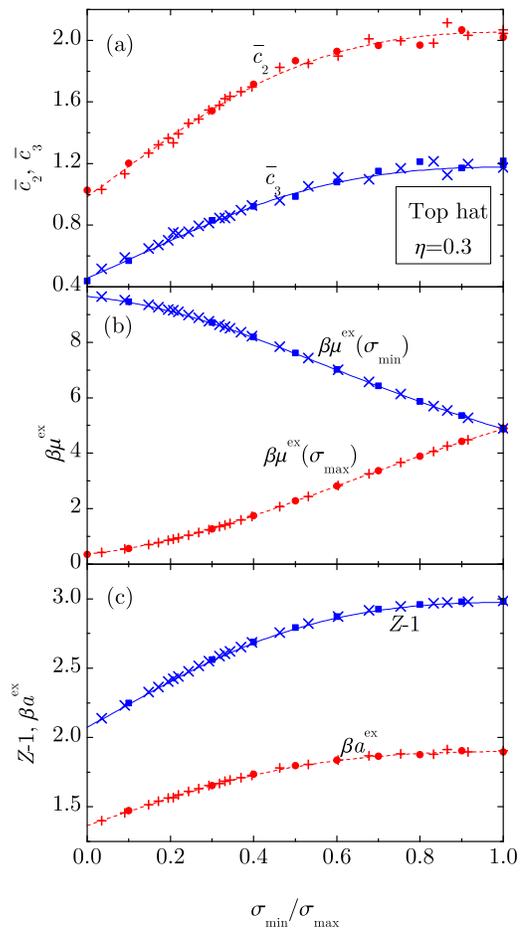}
\caption{Plot of (a) $\cc_2$ and $\cc_3$, (b) $\beta\mu^\ex(\sigma_{\min})$ and  $\beta\mu^\ex(\sigma_{\max})$, and (c) $Z-1$ and $\beta a^\ex$ versus $\sigma_{\min}/\sigma_{\max}$ for system TH1 (squares and circles). The crosses represent results obtained from system B1 by application of the mapping  \eqref{qTH_from_q}. The symbols are our MD data ($N=2048$) and the lines are the BMCSL predictions.\label{fig:TH1}}
\end{figure}

\begin{figure}[tbp]
\includegraphics[width=7cm]{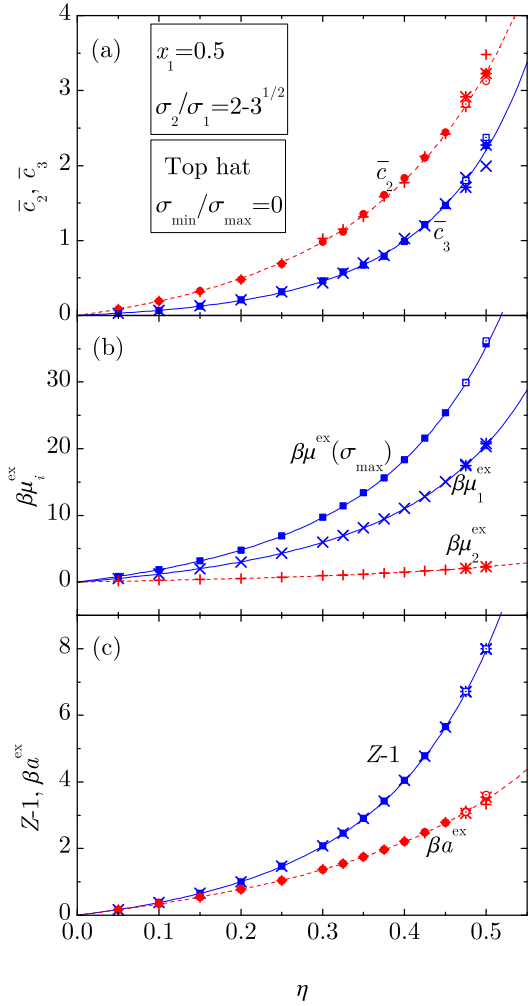}
\caption{Plot of (a) $\cc_2$ and $\cc_3$, (b) $\beta\mu_1^\ex$, $\beta\mu_2^\ex$, and  $\beta\mu^\ex(\sigma_{\max})$, and (c) $Z-1$ and $\beta a^\ex$ versus $\eta$ for systems B5 (crosses: $N=2048$, stars: $N=4000$) and TH2 (closed circles and squares: $N=2048$, open circles and squares: $N=4000$). The symbols are our MD data and the lines are the BMCSL predictions.\label{fig:B5&TH2}}
\end{figure}

\begin{figure*}[tbp]
\includegraphics[width=16cm]{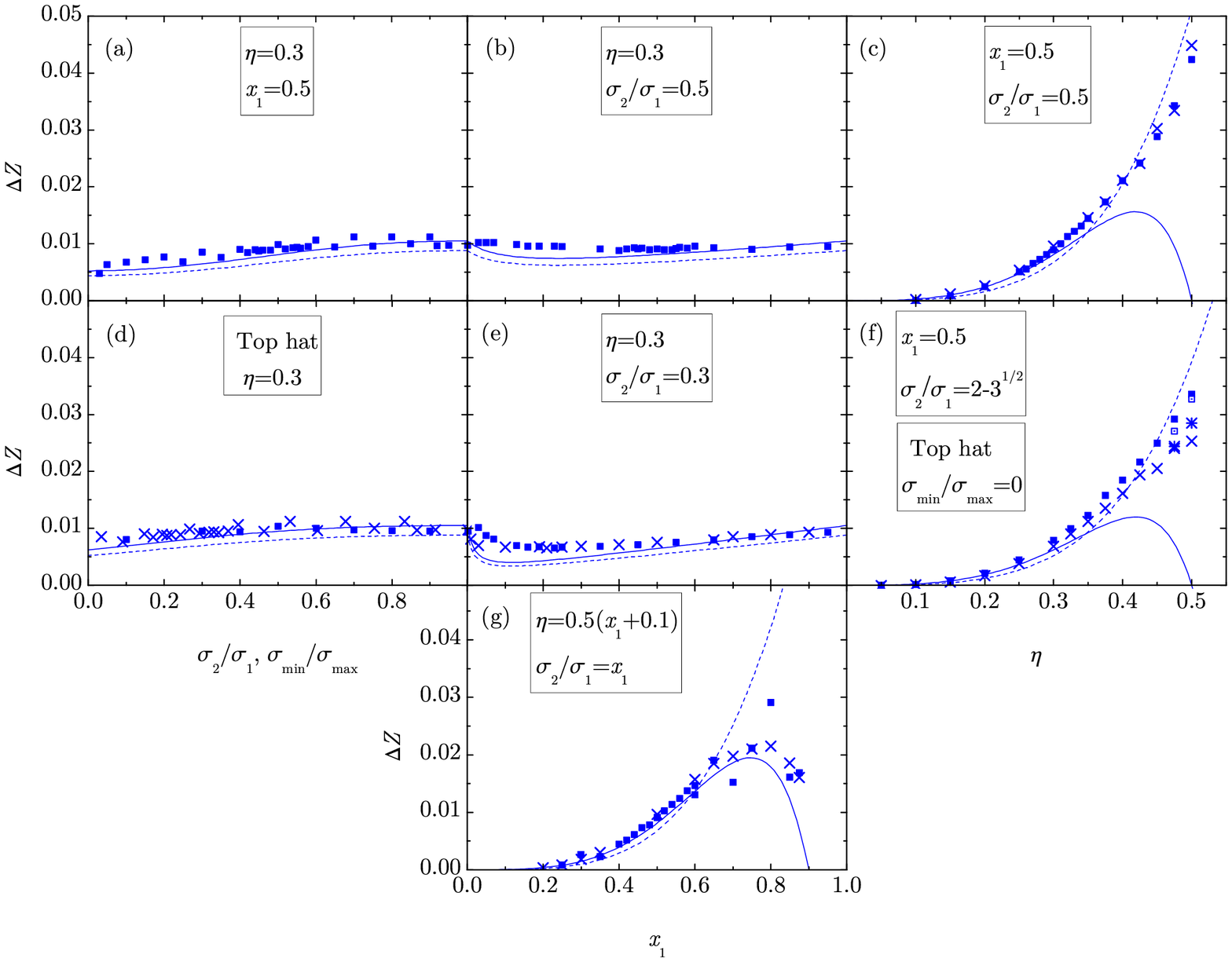}
\caption{Plot of the difference $\Delta Z=Z-Z^{\text{BMCSL}}$ for systems (a) B1, (b) B2, (c) B4 (squares: $N=2048$, crosses: $N=4000$), (d) TH1 (squares), (e) B3 (squares: $N=2048$, crosses: $N=4000$), (f) B5 (crosses: $N=2048$, stars: $N=4000$) and TH2 (closed squares: $N=2048$, open squares: $N=4000$), and (g) B6 (squares: $N=2048$, crosses: $N=4000$). The symbols are our MD data and the solid and dashed lines represent  $Z^{\text{BCSK}}-Z^{\text{BMCSL}}$ and $Z^{\text{PY-$\mu$c}}-Z^{\text{BMCSL}}$ (with $\alpha=0.37$), respectively. The crosses in panel (d) represent results obtained from system B1 by application of the mapping  \eqref{qTH_from_q}.\label{fig:DeltaZ}}
\end{figure*}

\section{Results and Discussion}
\label{sec3}

A number of binary and  polydisperse sets of mixtures were investigated for this study.
Cases were considered where a significant degree of ``separation'' from the one-component case ({see the Appendix})
would be expected.
Binary mixture simulations were carried out in which one of the three parameters (diameter ratio, $\sigma_{2}/\sigma_{1}$,  mole fraction  of the big spheres, $x_1$, and  total volume fraction, $\eta$) was varied, while the remaining two were kept constant.
An additional set was considered where  $\sigma_{2}/\sigma_{1}$ and $\eta$ changed linearly with $x_1$.
Also the TH particle distribution was  modeled for a range of $\sigma_{\min}/\sigma_{\max}$.   Details of the sets of
simulations carried out and the tags used for them are given in Table \ref{table3}.
Note that, according to theory and by application of Eq.\ \eqref{qTH}, the set of equimolar binary mixtures B1 and the set of polydisperse mixtures TH1 can be made equivalent. Likewise, the sets B5 and TH2 are theoretically equivalent.

The MD numerical values of $\cc_2$, $\cc_3$, $\beta\mu_1^\ex$, $\beta\mu_2^\ex$, $Z$, and $\beta a^\ex$ for the sets described in Table \ref{table3} are presented in Tables I-VIII of the supplementary material.

As for the different theoretical approaches summarized in Table \ref{table1}, we will discard the  PY-v, PY-c, and PY-$\mu$ EoSs since they are known to be clearly inferior to the BMCSL or BCSK prescriptions.\cite{SR13,HS16} Moreover,  the differences (both absolute and relative) between the BMCSL and BCSK predictions for the coefficients $\cc_2$ and $\cc_3$ are smaller than the (already very small) differences in the one-component case (where $\cc_2\to c_2$ and $\cc_3\to c_3$). This can be explained by the fact that $\cc_n^{\text{BMCSL}}-\cc_n^{\text{BCSK}}=\lambda^n\left(c_n^{\text{BMCSL}}-c_n^{\text{BCSK}}\right)$ and $\lambda<1$.
Therefore, in what follows we will mainly restrict ourselves to comparing the BMCSL EoS  with our MD results.

The simulation and theoretical results for the binary systems B1--B4 and B6 are presented in Figs.\ \ref{fig:B1}--\ref{fig:B4} and \ref{fig:B6}, respectively. Figure \ref{fig:TH1} does the same for the continuous distribution TH1 and again for the binary mixture B1 [in the latter case by applying the mapping \eqref{qTH_from_q}], while Fig.\ \ref{fig:B5&TH2} presents the results for systems B5 and TH2.

Even though the combined scope of these mixtures spans a wide spectrum of parameters,  very good
agreement
of the BMCSL theory with the MD results is observed in all the cases.
In particular, Figs.\ \ref{fig:TH1} and  \ref{fig:B5&TH2} confirm that systems as different as a binary mixture and a continuous size distribution can be practically indistinguishable from the thermodynamic point of view, provided they share the same values of $\lambda$ and $\gamma$. {Theory as well as simulation \cite{OL12} strongly support that this mapping property extends to other continuous distributions different from the TH one.}

Figures \ref{fig:B1}--\ref{fig:B5&TH2} show that, as expected, the fitting coefficients $\cc_2$ and $\cc_3$ present a certain  degree of scatter. However, there is some ``synergy'' or compensation between
these two quantities that takes place during the fitting process and reduces the level of scatter in the evaluation of
the excess chemical potentials and the free energy via Eqs.\ \eqref{2.11} and \eqref{3.1}, respectively. In any case, since the compressibility factor $Z$ is measured directly in the simulations (rather than by a fitting algorithm), it is a more robust quantity.
The excellent agreement found in Figs.\ \ref{fig:B3}--\ref{fig:B6}, and \ref{fig:B5&TH2} between the  $N=2048$ and $N=4000$ sets of MD data gives us confidence in the lack of any statistically significant finite-size effects.

Despite the good behavior of the BMCSL theory observed in Figs.\ \ref{fig:B1}--\ref{fig:B5&TH2}, a careful comparison shows that the theory generally underestimates the simulation values, especially in the case of the compressibility factor $Z$. The deviations $\Delta Z=Z-Z^{\text{BMCSL}}$ of the simulation data from the BMCSL predictions are plotted in Fig.\ \ref{fig:DeltaZ}, where also the deviations predicted by the BCSK EoS and the PY-$\mu$c EoS with $\alpha=0.37$ [see Eq.\ \eqref{PYmuc}] are also included.
We observe from Figs.\ \ref{fig:DeltaZ}(a), \ref{fig:DeltaZ}(b), \ref{fig:DeltaZ}(d), and \ref{fig:DeltaZ}(e)  that at a packing fraction $\eta=0.3$ one has $\Delta Z\approx 0.01$, practically with independence of the mixture composition. This property is very well accounted for by the PY-$\mu$c and, especially, BCSK theories. The same approximate value $\Delta Z\approx 0.01$ can be observed from Figs.\ \ref{fig:DeltaZ}(c) and \ref{fig:DeltaZ}(f) at $\eta=0.3$ and from Fig.\ \ref{fig:DeltaZ}(f) at $x_1=0.5$ (which implies $\eta=0.3$ in the set B6). As density increases in Figs.\ \ref{fig:DeltaZ}(c) and \ref{fig:DeltaZ}(f), $\Delta Z$ increases monotonically, reaching values close to $\Delta Z\approx 0.05$ at $\eta=0.5$. This trend is well captured by the PY-$\mu$c EoS, but not  by the BCSK EoS. The latter presents a maximum deviation $\Delta Z=(\lambda^3/\gamma)\left(223-70\sqrt{10}\right)/81\simeq 0.02 \lambda^3/\gamma$ at $\eta=2-\sqrt{5/2}\simeq 0.42$ and then $\Delta Z=0$ at $\eta=0.5$.
On the other hand, a similar non-monotonic behavior of $\Delta Z$ vs $x_1$ predicted by the BCSK EoS in Fig.\ \ref{fig:DeltaZ}(g) is actually confirmed by our simulations of the set B6, while the PY-$\mu$c EoS exhibits a {monotonic} behavior. Since  increasing the mole fraction $x_1$ in the set B6 implies approaching a monodisperse system at higher densities, we can conclude that, at packing fractions larger than about $\eta=0.35$, the BCSK EoS is more accurate than the PY-$\mu$c EoS for nearly monodisperse systems ($\lambda^3/\gamma\lesssim 1$), but the opposite happens for mixtures where $1-\lambda^3/\gamma$ is not small. For instance, $1-\lambda^3/\gamma\simeq 0.23$ and $0.41$ in Figs.\ \ref{fig:DeltaZ}(c) and \ref{fig:DeltaZ}(f), respectively, while $1-\lambda^3/\gamma<0.06$ for $x_1>0.7$ in Fig.\ \ref{fig:DeltaZ}(g).
Finally, it is interesting to notice from Fig.\ \ref{fig:DeltaZ}(f) a very slight breaking down of the equivalence between the mixtures B5 and TH2 as density increases.

\section{Conclusions}
\label{sec4}

An extensive series of sets of MD simulations were carried out of the
thermodynamic properties of binary and continuous size distribution
HS mixtures.  Relative particle diameters,
mole fractions of the different components, and the total packing (volume) fraction were systematically varied.
The Widom particle insertion method, employing  the Lab\'{i}k and
Smith technique,\cite{LS94} was used
to calculate the excess chemical potential
of a test particle of variable diameter, $\sigma_0$.
The simulation data
was fitted to a third-order polynomial in $\sigma_0$, the first two coefficients of which
are known exactly by theory. The compressibility factor, $Z$, was also independently obtained by a standard MD method.
As a novel outcome, the  excess free energy per particle was determined using a thermodynamic
relation involving the compressibility factor and the two fitted coefficients.  \\

The theories considered in this work share the same structure for the excess free energy and, hence, for the chemical potentials and the compressibility factor [see Eqs.\ \eqref{2.1}, \eqref{aex}, and \eqref{2.11}]. {These} theories differ only in the density dependence of the coefficient $a_2(\eta)$, since the coefficients $Z_2(\eta)$, $c_2(\eta)$, and $c_3(\eta)$ are derived from $a_2(\eta)$ by thermodynamic relations [see Eqs.\ \eqref{2.4b} and \eqref{2.14.1&2.14}]. The most widely used theory in the literature is the BMCSL, which is an interpolation between the virial and compressibility routes in the PY approximation. Here we have also taken the BCSK (an \emph{ad hoc} correction to the BMCSL EoS) and the PY$\mu$-c theories. The latter is an interpolation between the chemical-potential and compressibility routes in the PY approximation and thus it has the same footing as the BMCSL theory.

Very good agreement between the simulation results
with the predictions of the BMCSL analytic
EoS
is observed in all the cases. These simulations also confirm
that systems as different as a binary mixture and a continuous size distribution can be
hardly distinguishable in their thermodynamic quantities, provided they share the same values of
the parameters  $\lambda$ and $\gamma$, which mark the extent of
difference from the single component case.\\

A fine resolution examination of the MD generated quantities shows that the BMCSL theory
typically  underestimates the simulation values by a small amount,
especially for the compressibility factor.
These differences are generally captured well by the  BCSK and PY-$\mu$c  formulas (the latter with a mixing parameter $\alpha=0.37$) in different regions of the system parameter space.
When  the packing fraction is larger than about $\eta=0.35$, the PY-$\mu$c EoS is more accurate than the BCSK EoS, except for nearly monodisperse systems.\\

To conclude, we believe that the results reported here provide further evidence on the reliability of the BMCSL EoS over a wide spectrum of parameters characterizing a polydisperse HS fluid. On the other hand,  the BCSK and PY-$\mu$c EoSs, while formally similar to the BMCSL EoS, succeed in improving the theoretical predictions. We expect that the MD simulation data obtained in this work can be useful to test other alternative theories proposed in the literature.

\section*{Supplementary Material}
See supplementary material for  tables containing the MD simulation results for the mixtures described in Table \ref{table3}.

\begin{acknowledgments}
A.S. acknowledges the financial support of the
Ministerio de Econom\'ia y Competitividad (Spain) through Grant No.\ FIS2016-76359-P, partially financed by ``Fondo Europeo de Desarrollo Regional'' funds.
D.M.H. would like to thank Dr.\ T. Crane (Department of Physics, Royal Holloway, University of London, UK) for helpful software support.

\end{acknowledgments}

\appendix*
\begin{figure}[tbp]
\includegraphics[width=7cm]{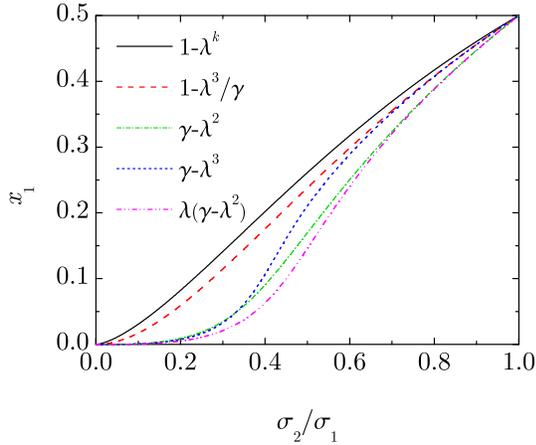}
\caption{{Dependence on $\sigma_2/\sigma_1$ of the mole fraction corresponding to the maximum values of $1-\lambda^k$, $1-\lambda^3/\gamma$, $\gamma-\lambda^2$, $\gamma-\lambda^3$, and $\lambda(\gamma-\lambda^2)$.}} \label{fig:x_max}
\end{figure}

{
\section{Extremal properties of combinations of $\lambda$ and $\gamma$ in binary mixtures}
\label{sec2.E}
In the framework of Eq.\ \eqref{2.1},  the deviations of the compressibility factor of a mixture  from that of the single component fluid (at a common value of the packing fraction $\eta$) are monitored by the (positive) differences $1-\lambda$ and $1-\lambda^3/\gamma$. In the case of a binary mixture characterized by the two parameters $x_1$ and $q= \sigma_2/\sigma_1\leq 1$, it can be checked that the maxima of $1-\lambda$ and $1-\lambda^3/\gamma$, at a fixed value of $q$, are
\begin{subequations}
\beq
\left(1-\lambda\right)_{\max}=\frac{(1-\sqrt{q})^2(1+q)}{(1-\sqrt{q}+q)^2},
\eeq
\beq
\left(1-\frac{\lambda^3}{\gamma}\right)_{\max}=
\frac{(1-q)^2 (2+q)^2 (1+2 q)^2}{4 \left(1+ q+q^2\right)^3}.
\eeq
\end{subequations}
Those maxima occur at
\begin{subequations}
\beq
\label{xL}
x_{1}=\left(q^{-3/2}+1\right)^{-1},
\eeq
\beq
x_1=\frac{q^2 (2+q)}{(1+q)\left(1+ q+q^2\right)},
\eeq
\end{subequations}
respectively.}

{
In the case of the chemical potential, we see from Eqs.\ \eqref{2.11} and \eqref{cc1-cc3} that the deviations are now measured by the differences $1-\lambda^k$ (with $k=1,2,3$), $\gamma-\lambda^2$, $\gamma-\lambda^3$, and $\lambda(\gamma-\lambda^2)$. In the case of $1-\lambda^k$, the maxima are located at \eqref{xL}, but the analytical expressions of the locations for the other quantities are too cumbersome to be reproduced here.}

{
Figure \ref{fig:x_max} shows the $q$-dependence of the mole fraction $x_1$ corresponding to the maxima of $1-\lambda^k$, $1-\lambda^3/\gamma$, $\gamma-\lambda^2$, $\gamma-\lambda^3$, and $\lambda(\gamma-\lambda^2)$.
In all the cases $x_1(q)$ decreases monotonically with decreasing $q$. In the limit $q\to 0$, while $x_1\sim q^{3/2}$ and $x_1\sim q^2$ in the cases of
$1-\lambda^k$ and $1-\lambda^3/\gamma$, respectively, the asymptotic behavior is $x_1\sim q^3$ in the cases of $\gamma-\lambda^2$, $\gamma-\lambda^3$, and $\lambda(\gamma-\lambda^2)$.}

\bibliography{D:/Dropbox/Mis_Dropcumentos/bib_files/liquid}

\begin{thebibliography}{51}%
\makeatletter
\providecommand \@ifxundefined [1]{%
 \@ifx{#1\undefined}
}%
\providecommand \@ifnum [1]{%
 \ifnum #1\expandafter \@firstoftwo
 \else \expandafter \@secondoftwo
 \fi
}%
\providecommand \@ifx [1]{%
 \ifx #1\expandafter \@firstoftwo
 \else \expandafter \@secondoftwo
 \fi
}%
\providecommand \natexlab [1]{#1}%
\providecommand \enquote  [1]{``#1''}%
\providecommand \bibnamefont  [1]{#1}%
\providecommand \bibfnamefont [1]{#1}%
\providecommand \citenamefont [1]{#1}%
\providecommand \href@noop [0]{\@secondoftwo}%
\providecommand \href [0]{\begingroup \@sanitize@url \@href}%
\providecommand \@href[1]{\@@startlink{#1}\@@href}%
\providecommand \@@href[1]{\endgroup#1\@@endlink}%
\providecommand \@sanitize@url [0]{\catcode `\\12\catcode `\$12\catcode
  `\&12\catcode `\#12\catcode `\^12\catcode `\_12\catcode `\%12\relax}%
\providecommand \@@startlink[1]{}%
\providecommand \@@endlink[0]{}%
\providecommand \url  [0]{\begingroup\@sanitize@url \@url }%
\providecommand \@url [1]{\endgroup\@href {#1}{\urlprefix }}%
\providecommand \urlprefix  [0]{URL }%
\providecommand \Eprint [0]{\href }%
\providecommand \doibase [0]{http://dx.doi.org/}%
\providecommand \selectlanguage [0]{\@gobble}%
\providecommand \bibinfo  [0]{\@secondoftwo}%
\providecommand \bibfield  [0]{\@secondoftwo}%
\providecommand \translation [1]{[#1]}%
\providecommand \BibitemOpen [0]{}%
\providecommand \bibitemStop [0]{}%
\providecommand \bibitemNoStop [0]{.\EOS\space}%
\providecommand \EOS [0]{\spacefactor3000\relax}%
\providecommand \BibitemShut  [1]{\csname bibitem#1\endcsname}%
\let\auto@bib@innerbib\@empty
%</preamble>
\bibitem [{\citenamefont {Boubl\'{i}k}(1970)}]{B70}%
  \BibitemOpen
  \bibfield  {author} {\bibinfo {author} {\bibfnamefont {T.}~\bibnamefont
  {Boubl\'{i}k}},\ }\bibfield  {title} {\enquote {\bibinfo {title} {Hard-sphere
  equation of state},}\ }\href {\doibase 10.1063/1.1673824} {\bibfield
  {journal} {\bibinfo  {journal} {J. Chem. Phys.}\ }\textbf {\bibinfo {volume}
  {53}},\ \bibinfo {pages} {471--472} (\bibinfo {year} {1970})}\BibitemShut
  {NoStop}%
\bibitem [{\citenamefont {Mansoori}\ \emph {et~al.}(1971)\citenamefont
  {Mansoori}, \citenamefont {Carnahan}, \citenamefont {Starling},\ and\
  \citenamefont {Leland{, Jr.}}}]{MCSL71}%
  \BibitemOpen
  \bibfield  {author} {\bibinfo {author} {\bibfnamefont {G.~A.}\ \bibnamefont
  {Mansoori}}, \bibinfo {author} {\bibfnamefont {N.~F.}\ \bibnamefont
  {Carnahan}}, \bibinfo {author} {\bibfnamefont {K.~E.}\ \bibnamefont
  {Starling}}, \ and\ \bibinfo {author} {\bibfnamefont {T.~W.}\ \bibnamefont
  {Leland{, Jr.}}},\ }\bibfield  {title} {\enquote {\bibinfo {title}
  {Equilibrium thermodynamic properties of the mixture of hard spheres},}\
  }\href {\doibase 10.1063/1.1675048} {\bibfield  {journal} {\bibinfo
  {journal} {J. Chem. Phys.}\ }\textbf {\bibinfo {volume} {54}},\ \bibinfo
  {pages} {1523--1525} (\bibinfo {year} {1971})}\BibitemShut {NoStop}%
\bibitem [{\citenamefont {Yau}, \citenamefont {Chan},\ and\ \citenamefont
  {Henderson}(1996)}]{YCH96}%
  \BibitemOpen
  \bibfield  {author} {\bibinfo {author} {\bibfnamefont {D.~H.~L.}\
  \bibnamefont {Yau}}, \bibinfo {author} {\bibfnamefont {K.-Y.}\ \bibnamefont
  {Chan}}, \ and\ \bibinfo {author} {\bibfnamefont {D.}~\bibnamefont
  {Henderson}},\ }\bibfield  {title} {\enquote {\bibinfo {title} {A further
  test of the {B}oublik et al. equations for binary hard sphere mixtures},}\
  }\href {\doibase 10.1080/00268979609484508} {\bibfield  {journal} {\bibinfo
  {journal} {Mol. Phys.}\ }\textbf {\bibinfo {volume} {88}},\ \bibinfo {pages}
  {1237--1248} (\bibinfo {year} {1996})}\BibitemShut {NoStop}%
\bibitem [{\citenamefont {Baro\v{s}ov\'a}\ \emph {et~al.}(1996)\citenamefont
  {Baro\v{s}ov\'a}, \citenamefont {Malijevsk\'y}, \citenamefont {Lab\'{\i}k},\
  and\ \citenamefont {Smith}}]{BMLS96}%
  \BibitemOpen
  \bibfield  {author} {\bibinfo {author} {\bibfnamefont {M.}~\bibnamefont
  {Baro\v{s}ov\'a}}, \bibinfo {author} {\bibfnamefont {A.}~\bibnamefont
  {Malijevsk\'y}}, \bibinfo {author} {\bibfnamefont {S.}~\bibnamefont
  {Lab\'{\i}k}}, \ and\ \bibinfo {author} {\bibfnamefont {W.~R.}\ \bibnamefont
  {Smith}},\ }\bibfield  {title} {\enquote {\bibinfo {title} {Computer
  simulation of the chemical potentials of binary hard-sphere mixtures},}\
  }\href {\doibase 10.1080/00268979600100281} {\bibfield  {journal} {\bibinfo
  {journal} {Mol. Phys.}\ }\textbf {\bibinfo {volume} {87}},\ \bibinfo {pages}
  {423--439} (\bibinfo {year} {1996})}\BibitemShut {NoStop}%
\bibitem [{\citenamefont {Liu}(1996)}]{L96}%
  \BibitemOpen
  \bibfield  {author} {\bibinfo {author} {\bibfnamefont {A.}~\bibnamefont
  {Liu}},\ }\bibfield  {title} {\enquote {\bibinfo {title} {Calculation of
  chemical potentials of mixtures from computer simulations},}\ }\href
  {\doibase 10.1080/002689796172986} {\bibfield  {journal} {\bibinfo  {journal}
  {Mol. Phys.}\ }\textbf {\bibinfo {volume} {89}},\ \bibinfo {pages}
  {1651--1658} (\bibinfo {year} {1996})}\BibitemShut {NoStop}%
\bibitem [{\citenamefont {Lue}\ and\ \citenamefont {Woodcock}(1999)}]{LW99}%
  \BibitemOpen
  \bibfield  {author} {\bibinfo {author} {\bibfnamefont {L.}~\bibnamefont
  {Lue}}\ and\ \bibinfo {author} {\bibfnamefont {L.~V.}\ \bibnamefont
  {Woodcock}},\ }\bibfield  {title} {\enquote {\bibinfo {title} {{Depletion
  effects and gelation in a binary hard-sphere fluid}},}\ }\href {\doibase
  10.1080/00268979909483087} {\bibfield  {journal} {\bibinfo  {journal} {Mol.
  Phys.}\ }\textbf {\bibinfo {volume} {96}},\ \bibinfo {pages} {1435--1443}
  (\bibinfo {year} {1999})}\BibitemShut {NoStop}%
\bibitem [{\citenamefont {Tokuyama}\ and\ \citenamefont {Terada}(2005)}]{TT05}%
  \BibitemOpen
  \bibfield  {author} {\bibinfo {author} {\bibfnamefont {M.}~\bibnamefont
  {Tokuyama}}\ and\ \bibinfo {author} {\bibfnamefont {Y.}~\bibnamefont
  {Terada}},\ }\bibfield  {title} {\enquote {\bibinfo {title} {Slow dynamics
  and re-entrant melting in a polydisperse hard-sphere fluid},}\ }\href
  {\doibase 10.1021%2Fjp0544383} {\bibfield  {journal} {\bibinfo  {journal} {J.
  Phys. Chem. B}\ }\textbf {\bibinfo {volume} {109}},\ \bibinfo {pages}
  {21357--21363} (\bibinfo {year} {2005})}\BibitemShut {NoStop}%
\bibitem [{\citenamefont {Barrio}\ and\ \citenamefont {Solana}(2008)}]{BS08}%
  \BibitemOpen
  \bibfield  {author} {\bibinfo {author} {\bibfnamefont {C.}~\bibnamefont
  {Barrio}}\ and\ \bibinfo {author} {\bibfnamefont {J.~R.}\ \bibnamefont
  {Solana}},\ }\bibfield  {title} {\enquote {\bibinfo {title} {{Binary Mixtures
  of Additive Hard Spheres. Simulations and Theories}},}\ }in\ \href@noop {}
  {\emph {\bibinfo {booktitle} {{Theory and Simulation of Hard-Sphere Fluids
  and Related Systems}}}},\ \bibinfo {series} {Lecture Notes in Physics}, Vol.\
  \bibinfo {volume} {753},\ \bibinfo {editor} {edited by\ \bibinfo {editor}
  {\bibfnamefont {A.}~\bibnamefont {Mulero}}}\ (\bibinfo  {publisher}
  {Springer-Verlag},\ \bibinfo {address} {Berlin},\ \bibinfo {year} {2008})\
  pp.\ \bibinfo {pages} {133--182}\BibitemShut {NoStop}%
\bibitem [{\citenamefont {Odriozola}\ and\ \citenamefont
  {Berthier}(2011)}]{OB11}%
  \BibitemOpen
  \bibfield  {author} {\bibinfo {author} {\bibfnamefont {G.}~\bibnamefont
  {Odriozola}}\ and\ \bibinfo {author} {\bibfnamefont {L.}~\bibnamefont
  {Berthier}},\ }\bibfield  {title} {\enquote {\bibinfo {title} {Equilibrium
  equation of state of a hard sphere binary mixture at very large densities
  using replica exchange {M}onte {C}arlo simulations},}\ }\href {\doibase
  10.1063/1.3541248} {\bibfield  {journal} {\bibinfo  {journal} {J. Chem.
  Phys.}\ }\textbf {\bibinfo {volume} {134}},\ \bibinfo {pages} {054504}
  (\bibinfo {year} {2011})}\BibitemShut {NoStop}%
\bibitem [{\citenamefont {Widom}(1963)}]{W63b}%
  \BibitemOpen
  \bibfield  {author} {\bibinfo {author} {\bibfnamefont {B.}~\bibnamefont
  {Widom}},\ }\bibfield  {title} {\enquote {\bibinfo {title} {Some topics in
  the theory of fluids},}\ }\href {\doibase 10.1063/1.1734110} {\bibfield
  {journal} {\bibinfo  {journal} {J. Chem. Phys.}\ }\textbf {\bibinfo {volume}
  {39}},\ \bibinfo {pages} {2808--2812} (\bibinfo {year} {1963})}\BibitemShut
  {NoStop}%
\bibitem [{\citenamefont {Lab\'ik}\ and\ \citenamefont {Smith}(1994)}]{LS94}%
  \BibitemOpen
  \bibfield  {author} {\bibinfo {author} {\bibfnamefont {S.}~\bibnamefont
  {Lab\'ik}}\ and\ \bibinfo {author} {\bibfnamefont {W.~R.}\ \bibnamefont
  {Smith}},\ }\bibfield  {title} {\enquote {\bibinfo {title} {Scaled {P}article
  {T}heory and the efficient calculation of the chemical potential of hard
  spheres in the {NVT} ensemble},}\ }\href {\doibase 10.1080/08927029408022533}
  {\bibfield  {journal} {\bibinfo  {journal} {Mol. Simul.}\ }\textbf {\bibinfo
  {volume} {12}},\ \bibinfo {pages} {23--31} (\bibinfo {year}
  {1994})}\BibitemShut {NoStop}%
\bibitem [{\citenamefont {Heyes}\ and\ \citenamefont {Santos}(2016)}]{HS16}%
  \BibitemOpen
  \bibfield  {author} {\bibinfo {author} {\bibfnamefont {D.~M.}\ \bibnamefont
  {Heyes}}\ and\ \bibinfo {author} {\bibfnamefont {A.}~\bibnamefont {Santos}},\
  }\bibfield  {title} {\enquote {\bibinfo {title} {Chemical potential of a test
  hard sphere of variable size in a hard-sphere fluid},}\ }\href {\doibase
  10.1063/1.4968039} {\bibfield  {journal} {\bibinfo  {journal} {J. Chem.
  Phys.}\ }\textbf {\bibinfo {volume} {145}},\ \bibinfo {pages} {214504}
  (\bibinfo {year} {2016})}\BibitemShut {NoStop}%
\bibitem [{\citenamefont {Baranau}\ and\ \citenamefont
  {Tallarek}(2016)}]{BT16}%
  \BibitemOpen
  \bibfield  {author} {\bibinfo {author} {\bibfnamefont {V.}~\bibnamefont
  {Baranau}}\ and\ \bibinfo {author} {\bibfnamefont {U.}~\bibnamefont
  {Tallarek}},\ }\bibfield  {title} {\enquote {\bibinfo {title} {Chemical
  potential and entropy in monodisperse and polydisperse hard-sphere fluids
  using {W}idom's particle insertion method and a pore size distribution-based
  insertion probability},}\ }\href {\doibase 10.1063/1.4953079} {\bibfield
  {journal} {\bibinfo  {journal} {J. Chem. Phys.}\ }\textbf {\bibinfo {volume}
  {144}},\ \bibinfo {pages} {214503} (\bibinfo {year} {2016})}\BibitemShut
  {NoStop}%
\bibitem [{\citenamefont {Heyes}(1992)}]{H92}%
  \BibitemOpen
  \bibfield  {author} {\bibinfo {author} {\bibfnamefont {D.~M.}\ \bibnamefont
  {Heyes}},\ }\bibfield  {title} {\enquote {\bibinfo {title} {Chemical
  potential, partial enthalpy and partial volume of mixtures by {NPT} molecular
  dynamics},}\ }\href {\doibase 10.1016/0301-0104(92)80067-6} {\bibfield
  {journal} {\bibinfo  {journal} {Chem. Phys.}\ }\textbf {\bibinfo {volume}
  {159}},\ \bibinfo {pages} {149--167} (\bibinfo {year} {1992})}\BibitemShut
  {NoStop}%
\bibitem [{\citenamefont {Perego}, \citenamefont {Giberti},\ and\ \citenamefont
  {Parrinello}(2016)}]{PGP16}%
  \BibitemOpen
  \bibfield  {author} {\bibinfo {author} {\bibfnamefont {C.}~\bibnamefont
  {Perego}}, \bibinfo {author} {\bibfnamefont {F.}~\bibnamefont {Giberti}}, \
  and\ \bibinfo {author} {\bibfnamefont {M.}~\bibnamefont {Parrinello}},\
  }\bibfield  {title} {\enquote {\bibinfo {title} {Chemical potential
  calculations in dense liquids using metadynamics},}\ }\href {\doibase
  10.1140/epjst/e2016-60094-x} {\bibfield  {journal} {\bibinfo  {journal} {Eur.
  Phys. J. Spec. Top.}\ }\textbf {\bibinfo {volume} {225}},\ \bibinfo {pages}
  {1621--1628} (\bibinfo {year} {2016})}\BibitemShut {NoStop}%
\bibitem [{\citenamefont {Valsson}, \citenamefont {Tiwary},\ and\ \citenamefont
  {Parrinello}(2016)}]{VTP16}%
  \BibitemOpen
  \bibfield  {author} {\bibinfo {author} {\bibfnamefont {O.}~\bibnamefont
  {Valsson}}, \bibinfo {author} {\bibfnamefont {P.}~\bibnamefont {Tiwary}}, \
  and\ \bibinfo {author} {\bibfnamefont {M.}~\bibnamefont {Parrinello}},\
  }\bibfield  {title} {\enquote {\bibinfo {title} {Enhancing important
  fluctuations: Rare events and metadynamics from a conceptual viewpoint},}\
  }\href {\doibase 10.1146/annurev-physchem-040215-112229} {\bibfield
  {journal} {\bibinfo  {journal} {Annu. Rev. Phys. Chem.}\ }\textbf {\bibinfo
  {volume} {67}},\ \bibinfo {pages} {159--184} (\bibinfo {year}
  {2016})}\BibitemShut {NoStop}%
\bibitem [{\citenamefont {Santos}, \citenamefont {Yuste},\ and\ \citenamefont
  {{L\'opez de Haro}}(1999)}]{SYH99}%
  \BibitemOpen
  \bibfield  {author} {\bibinfo {author} {\bibfnamefont {A.}~\bibnamefont
  {Santos}}, \bibinfo {author} {\bibfnamefont {S.~B.}\ \bibnamefont {Yuste}}, \
  and\ \bibinfo {author} {\bibfnamefont {M.}~\bibnamefont {{L\'opez de
  Haro}}},\ }\bibfield  {title} {\enquote {\bibinfo {title} {{Equation of state
  of a multicomponent $d$-dimensional hard-sphere fluid}},}\ }\href {\doibase
  10.1080/00268979909482932} {\bibfield  {journal} {\bibinfo  {journal} {Mol.
  Phys.}\ }\textbf {\bibinfo {volume} {96}},\ \bibinfo {pages} {1--5} (\bibinfo
  {year} {1999})}\BibitemShut {NoStop}%
\bibitem [{\citenamefont {Santos}, \citenamefont {Yuste},\ and\ \citenamefont
  {{L\'opez de Haro}}(2001)}]{SYH01}%
  \BibitemOpen
  \bibfield  {author} {\bibinfo {author} {\bibfnamefont {A.}~\bibnamefont
  {Santos}}, \bibinfo {author} {\bibfnamefont {S.~B.}\ \bibnamefont {Yuste}}, \
  and\ \bibinfo {author} {\bibfnamefont {M.}~\bibnamefont {{L\'opez de
  Haro}}},\ }\bibfield  {title} {\enquote {\bibinfo {title} {Virial
  coefficients and equations of state for mixtures of hard discs, hard spheres,
  and hard hyperspheres},}\ }\href {\doibase 10.1080/00268970110063890}
  {\bibfield  {journal} {\bibinfo  {journal} {Mol. Phys.}\ }\textbf {\bibinfo
  {volume} {99}},\ \bibinfo {pages} {1959--1972} (\bibinfo {year}
  {2001})}\BibitemShut {NoStop}%
\bibitem [{\citenamefont {Santos}, \citenamefont {Yuste},\ and\ \citenamefont
  {{L\'opez de Haro}}(2002)}]{SYH02}%
  \BibitemOpen
  \bibfield  {author} {\bibinfo {author} {\bibfnamefont {A.}~\bibnamefont
  {Santos}}, \bibinfo {author} {\bibfnamefont {S.~B.}\ \bibnamefont {Yuste}}, \
  and\ \bibinfo {author} {\bibfnamefont {M.}~\bibnamefont {{L\'opez de
  Haro}}},\ }\bibfield  {title} {\enquote {\bibinfo {title} {Contact values of
  the radial distribution functions of additive hard-sphere mixtures in $d$
  dimensions: A new proposal},}\ }\href {\doibase 10.1063/1.1502247} {\bibfield
   {journal} {\bibinfo  {journal} {J. Chem. Phys.}\ }\textbf {\bibinfo {volume}
  {117}},\ \bibinfo {pages} {5785--5793} (\bibinfo {year} {2002})}\BibitemShut
  {NoStop}%
\bibitem [{\citenamefont {Santos}, \citenamefont {Yuste},\ and\ \citenamefont
  {{L\'opez de Haro}}(2005)}]{SYH05}%
  \BibitemOpen
  \bibfield  {author} {\bibinfo {author} {\bibfnamefont {A.}~\bibnamefont
  {Santos}}, \bibinfo {author} {\bibfnamefont {S.~B.}\ \bibnamefont {Yuste}}, \
  and\ \bibinfo {author} {\bibfnamefont {M.}~\bibnamefont {{L\'opez de
  Haro}}},\ }\bibfield  {title} {\enquote {\bibinfo {title} {Contact values of
  the particle-particle and wall-particle correlation functions in a
  hard-sphere polydisperse fluid},}\ }\href {\doibase 10.1063/1.2136883}
  {\bibfield  {journal} {\bibinfo  {journal} {J. Chem. Phys.}\ }\textbf
  {\bibinfo {volume} {123}},\ \bibinfo {pages} {234512} (\bibinfo {year}
  {2005})}\BibitemShut {NoStop}%
\bibitem [{\citenamefont {Hansen-Goos}\ and\ \citenamefont
  {Roth}(2006)}]{HR06a}%
  \BibitemOpen
  \bibfield  {author} {\bibinfo {author} {\bibfnamefont {H.}~\bibnamefont
  {Hansen-Goos}}\ and\ \bibinfo {author} {\bibfnamefont {R.}~\bibnamefont
  {Roth}},\ }\bibfield  {title} {\enquote {\bibinfo {title} {A new
  generalization of the {Carnahan-Starling} equation of state to additive
  mixtures of hard spheres},}\ }\href {\doibase 10.1063/1.2187491} {\bibfield
  {journal} {\bibinfo  {journal} {J. Chem. Phys.}\ }\textbf {\bibinfo {volume}
  {124}},\ \bibinfo {pages} {154506} (\bibinfo {year} {2006})}\BibitemShut
  {NoStop}%
\bibitem [{\citenamefont {Paricaud}(2015)}]{P15}%
  \BibitemOpen
  \bibfield  {author} {\bibinfo {author} {\bibfnamefont {P.}~\bibnamefont
  {Paricaud}},\ }\bibfield  {title} {\enquote {\bibinfo {title} {Extension of
  the {BMCSL} equation of state for hard spheres to the metastable disordered
  region: Application to the {SAFT} approach},}\ }\href {\doibase
  10.1063/1.4927148} {\bibfield  {journal} {\bibinfo  {journal} {J. Chem.
  Phys.}\ }\textbf {\bibinfo {volume} {143}},\ \bibinfo {pages} {044507}
  (\bibinfo {year} {2015})}\BibitemShut {NoStop}%
\bibitem [{\citenamefont {Mulero}\ \emph {et~al.}(2008)\citenamefont {Mulero},
  \citenamefont {Gal\'an}, \citenamefont {Parra},\ and\ \citenamefont
  {Cuadros}}]{MGPC08}%
  \BibitemOpen
  \bibfield  {author} {\bibinfo {author} {\bibfnamefont {A.}~\bibnamefont
  {Mulero}}, \bibinfo {author} {\bibfnamefont {C.~A.}\ \bibnamefont {Gal\'an}},
  \bibinfo {author} {\bibfnamefont {M.~I.}\ \bibnamefont {Parra}}, \ and\
  \bibinfo {author} {\bibfnamefont {F.}~\bibnamefont {Cuadros}},\ }\bibfield
  {title} {\enquote {\bibinfo {title} {{Equations of State for Hard Spheres and
  Hard Disks}},}\ }in\ \href@noop {} {\emph {\bibinfo {booktitle} {{Theory and
  Simulation of Hard-Sphere Fluids and Related Systems}}}},\ \bibinfo {series}
  {Lecture Notes in Physics}, Vol.\ \bibinfo {volume} {753},\ \bibinfo {editor}
  {edited by\ \bibinfo {editor} {\bibfnamefont {A.}~\bibnamefont {Mulero}}}\
  (\bibinfo  {publisher} {Springer-Verlag},\ \bibinfo {address} {Berlin},\
  \bibinfo {year} {2008})\ pp.\ \bibinfo {pages} {37--109}\BibitemShut
  {NoStop}%
\bibitem [{\citenamefont {Lebowitz}\ and\ \citenamefont {Zomick}(1971)}]{LZ71}%
  \BibitemOpen
  \bibfield  {author} {\bibinfo {author} {\bibfnamefont {J.~L.}\ \bibnamefont
  {Lebowitz}}\ and\ \bibinfo {author} {\bibfnamefont {D.}~\bibnamefont
  {Zomick}},\ }\bibfield  {title} {\enquote {\bibinfo {title} {Mixtures of hard
  spheres with nonadditive diameters: {S}ome exact results and solution of {PY}
  equation},}\ }\href {\doibase 10.1063/1.1675348} {\bibfield  {journal}
  {\bibinfo  {journal} {J. Chem. Phys.}\ }\textbf {\bibinfo {volume} {54}},\
  \bibinfo {pages} {3335--3346} (\bibinfo {year} {1971})}\BibitemShut {NoStop}%
\bibitem [{\citenamefont {Perram}\ and\ \citenamefont {Smith}(1975)}]{PS75}%
  \BibitemOpen
  \bibfield  {author} {\bibinfo {author} {\bibfnamefont {J.~W.}\ \bibnamefont
  {Perram}}\ and\ \bibinfo {author} {\bibfnamefont {E.~R.}\ \bibnamefont
  {Smith}},\ }\bibfield  {title} {\enquote {\bibinfo {title} {A model for the
  examination of phase behaviour in multicomponent systems},}\ }\href {\doibase
  10.1016/0009-2614(75)85604-1} {\bibfield  {journal} {\bibinfo  {journal}
  {Chem. Phys. Lett.}\ }\textbf {\bibinfo {volume} {35}},\ \bibinfo {pages}
  {138--140} (\bibinfo {year} {1975})}\BibitemShut {NoStop}%
\bibitem [{\citenamefont {Barboy}(1975)}]{B75}%
  \BibitemOpen
  \bibfield  {author} {\bibinfo {author} {\bibfnamefont {B.}~\bibnamefont
  {Barboy}},\ }\bibfield  {title} {\enquote {\bibinfo {title} {Solution of the
  compressibility equation of the adhesive hard-sphere model for mixtures},}\
  }\href {\doibase 10.1016/0301-0104(75)80055-3} {\bibfield  {journal}
  {\bibinfo  {journal} {Chem. Phys.}\ }\textbf {\bibinfo {volume} {11}},\
  \bibinfo {pages} {357--371} (\bibinfo {year} {1975})}\BibitemShut {NoStop}%
\bibitem [{\citenamefont {Percus}\ and\ \citenamefont {Yevick}(1958)}]{PY58}%
  \BibitemOpen
  \bibfield  {author} {\bibinfo {author} {\bibfnamefont {J.~K.}\ \bibnamefont
  {Percus}}\ and\ \bibinfo {author} {\bibfnamefont {G.~J.}\ \bibnamefont
  {Yevick}},\ }\bibfield  {title} {\enquote {\bibinfo {title} {Analysis of
  classical statistical mechanics by means of collective coordinates},}\ }\href
  {\doibase 10.1103/PhysRev.110.1} {\bibfield  {journal} {\bibinfo  {journal}
  {Phys. Rev.}\ }\textbf {\bibinfo {volume} {110}},\ \bibinfo {pages} {1--13}
  (\bibinfo {year} {1958})}\BibitemShut {NoStop}%
\bibitem [{\citenamefont {Santos}(2016)}]{S16}%
  \BibitemOpen
  \bibfield  {author} {\bibinfo {author} {\bibfnamefont {A.}~\bibnamefont
  {Santos}},\ }\href@noop {} {\emph {\bibinfo {title} {{A Concise Course on the
  Theory of Classical Liquids. Basics and Selected Topics}}}},\ \bibinfo
  {series} {Lecture Notes in Physics}, Vol.\ \bibinfo {volume} {923}\ (\bibinfo
   {publisher} {Springer},\ \bibinfo {address} {New York},\ \bibinfo {year}
  {2016})\BibitemShut {NoStop}%
\bibitem [{\citenamefont {Santos}(2012{\natexlab{a}})}]{S12b}%
  \BibitemOpen
  \bibfield  {author} {\bibinfo {author} {\bibfnamefont {A.}~\bibnamefont
  {Santos}},\ }\bibfield  {title} {\enquote {\bibinfo {title}
  {Chemical-potential route: {A} hidden {P}ercus--{Y}evick equation of state
  for hard spheres},}\ }\href {\doibase 10.1103/PhysRevLett.109.120601}
  {\bibfield  {journal} {\bibinfo  {journal} {Phys. Rev. Lett.}\ }\textbf
  {\bibinfo {volume} {109}},\ \bibinfo {pages} {{120}{601}} (\bibinfo {year}
  {2012}{\natexlab{a}})}\BibitemShut {NoStop}%
\bibitem [{\citenamefont {Santos}\ and\ \citenamefont {Rohrmann}(2013)}]{SR13}%
  \BibitemOpen
  \bibfield  {author} {\bibinfo {author} {\bibfnamefont {A.}~\bibnamefont
  {Santos}}\ and\ \bibinfo {author} {\bibfnamefont {R.~D.}\ \bibnamefont
  {Rohrmann}},\ }\bibfield  {title} {\enquote {\bibinfo {title}
  {Chemical-potential route for multicomponent fluids},}\ }\href {\doibase
  10.1103/PhysRevE.87.052138} {\bibfield  {journal} {\bibinfo  {journal} {Phys.
  Rev. E}\ }\textbf {\bibinfo {volume} {87}},\ \bibinfo {pages} {{052}{138}}
  (\bibinfo {year} {2013})}\BibitemShut {NoStop}%
\bibitem [{\citenamefont {Ogarko}\ and\ \citenamefont {Luding}(2012)}]{OL12}%
  \BibitemOpen
  \bibfield  {author} {\bibinfo {author} {\bibfnamefont {V.}~\bibnamefont
  {Ogarko}}\ and\ \bibinfo {author} {\bibfnamefont {S.}~\bibnamefont
  {Luding}},\ }\bibfield  {title} {\enquote {\bibinfo {title} {Equation of
  state and jamming density for equivalent bi- and polydisperse, smooth, hard
  sphere systems},}\ }\href {\doibase 10.1063/1.3694030} {\bibfield  {journal}
  {\bibinfo  {journal} {J. Chem. Phys.}\ }\textbf {\bibinfo {volume} {136}},\
  \bibinfo {pages} {{124}{508}} (\bibinfo {year} {2012})}\BibitemShut {NoStop}%
\bibitem [{\citenamefont {Reiss}, \citenamefont {Frisch},\ and\ \citenamefont
  {Lebowitz}(1959)}]{RFL59}%
  \BibitemOpen
  \bibfield  {author} {\bibinfo {author} {\bibfnamefont {H.}~\bibnamefont
  {Reiss}}, \bibinfo {author} {\bibfnamefont {H.~L.}\ \bibnamefont {Frisch}}, \
  and\ \bibinfo {author} {\bibfnamefont {J.~L.}\ \bibnamefont {Lebowitz}},\
  }\bibfield  {title} {\enquote {\bibinfo {title} {Statistical mechanics of
  rigid spheres},}\ }\href {\doibase 10.1063/1.1730361} {\bibfield  {journal}
  {\bibinfo  {journal} {J. Chem. Phys.}\ }\textbf {\bibinfo {volume} {31}},\
  \bibinfo {pages} {369--380} (\bibinfo {year} {1959})}\BibitemShut {NoStop}%
\bibitem [{\citenamefont {Lebowitz}, \citenamefont {Helfand},\ and\
  \citenamefont {Praestgaard}(1965)}]{LHP65}%
  \BibitemOpen
  \bibfield  {author} {\bibinfo {author} {\bibfnamefont {J.~L.}\ \bibnamefont
  {Lebowitz}}, \bibinfo {author} {\bibfnamefont {E.}~\bibnamefont {Helfand}}, \
  and\ \bibinfo {author} {\bibfnamefont {E.}~\bibnamefont {Praestgaard}},\
  }\bibfield  {title} {\enquote {\bibinfo {title} {Scaled {P}article {T}heory
  of fluid mixtures},}\ }\href {\doibase 10.1063/1.1696842} {\bibfield
  {journal} {\bibinfo  {journal} {J. Chem. Phys.}\ }\textbf {\bibinfo {volume}
  {43}},\ \bibinfo {pages} {774--779} (\bibinfo {year} {1965})}\BibitemShut
  {NoStop}%
\bibitem [{\citenamefont {Mandell}\ and\ \citenamefont {Reiss}(1975)}]{MR75}%
  \BibitemOpen
  \bibfield  {author} {\bibinfo {author} {\bibfnamefont {M.}~\bibnamefont
  {Mandell}}\ and\ \bibinfo {author} {\bibfnamefont {H.}~\bibnamefont
  {Reiss}},\ }\bibfield  {title} {\enquote {\bibinfo {title} {Scaled {P}article
  {T}heory: {Solution} to the complete set of {S}caled {P}article {T}heory
  conditions: {Applications} to surface structure and dilute mixtures},}\
  }\href {\doibase 10.1007/BF01221372} {\bibfield  {journal} {\bibinfo
  {journal} {J. Stat. Phys.}\ }\textbf {\bibinfo {volume} {13}},\ \bibinfo
  {pages} {113--128} (\bibinfo {year} {1975})}\BibitemShut {NoStop}%
\bibitem [{\citenamefont {Rosenfeld}(1988)}]{R88}%
  \BibitemOpen
  \bibfield  {author} {\bibinfo {author} {\bibfnamefont {Y.}~\bibnamefont
  {Rosenfeld}},\ }\bibfield  {title} {\enquote {\bibinfo {title} {Scaled field
  particle theory of the structure and thermodynamics of isotropic hard
  particle fluids},}\ }\href {\doibase 10.1063/1.454810} {\bibfield  {journal}
  {\bibinfo  {journal} {J. Chem. Phys.}\ }\textbf {\bibinfo {volume} {89}},\
  \bibinfo {pages} {4272--4287} (\bibinfo {year} {1988})}\BibitemShut {NoStop}%
\bibitem [{\citenamefont {Heying}\ and\ \citenamefont {Corti}(2004)}]{HC04b}%
  \BibitemOpen
  \bibfield  {author} {\bibinfo {author} {\bibfnamefont {M.}~\bibnamefont
  {Heying}}\ and\ \bibinfo {author} {\bibfnamefont {D.}~\bibnamefont {Corti}},\
  }\bibfield  {title} {\enquote {\bibinfo {title} {Scaled {P}article {T}heory
  revisited: New conditions and improved predictions of the properties of the
  hard sphere fluid},}\ }\href {\doibase 10.1021/jp040398b} {\bibfield
  {journal} {\bibinfo  {journal} {J. Phys. Chem. B}\ }\textbf {\bibinfo
  {volume} {108}},\ \bibinfo {pages} {{19}{756}--{19}{768}} (\bibinfo {year}
  {2004})}\BibitemShut {NoStop}%
\bibitem [{\citenamefont {{L\'opez de Haro}}, \citenamefont {Yuste},\ and\
  \citenamefont {Santos}(2008)}]{HYS08}%
  \BibitemOpen
  \bibfield  {author} {\bibinfo {author} {\bibfnamefont {M.}~\bibnamefont
  {{L\'opez de Haro}}}, \bibinfo {author} {\bibfnamefont {S.~B.}\ \bibnamefont
  {Yuste}}, \ and\ \bibinfo {author} {\bibfnamefont {A.}~\bibnamefont
  {Santos}},\ }\bibfield  {title} {\enquote {\bibinfo {title} {{Alternative
  Approaches to the Equilibrium Properties of Hard-Sphere Liquids}},}\ }in\
  \href@noop {} {\emph {\bibinfo {booktitle} {{Theory and Simulation of
  Hard-Sphere Fluids and Related Systems}}}},\ \bibinfo {series} {Lecture Notes
  in Physics}, Vol.\ \bibinfo {volume} {753},\ \bibinfo {editor} {edited by\
  \bibinfo {editor} {\bibfnamefont {A.}~\bibnamefont {Mulero}}}\ (\bibinfo
  {publisher} {Springer},\ \bibinfo {address} {Berlin},\ \bibinfo {year}
  {2008})\ pp.\ \bibinfo {pages} {183--245}\BibitemShut {NoStop}%
\bibitem [{\citenamefont {Santos}, \citenamefont {Yuste},\ and\ \citenamefont
  {{L\'opez de Haro}}(2011)}]{SYH11}%
  \BibitemOpen
  \bibfield  {author} {\bibinfo {author} {\bibfnamefont {A.}~\bibnamefont
  {Santos}}, \bibinfo {author} {\bibfnamefont {S.~B.}\ \bibnamefont {Yuste}}, \
  and\ \bibinfo {author} {\bibfnamefont {M.}~\bibnamefont {{L\'opez de
  Haro}}},\ }\bibfield  {title} {\enquote {\bibinfo {title} {Communication:
  Inferring the equation of state of a metastable hard-sphere fluid from the
  equation of state of a hard-sphere mixture at high densities},}\ }\href
  {\doibase 10.1063/1.3663206} {\bibfield  {journal} {\bibinfo  {journal} {J.
  Chem. Phys.}\ }\textbf {\bibinfo {volume} {135}},\ \bibinfo {pages} {181102}
  (\bibinfo {year} {2011})}\BibitemShut {NoStop}%
\bibitem [{\citenamefont {Santos}(2012{\natexlab{b}})}]{S12c}%
  \BibitemOpen
  \bibfield  {author} {\bibinfo {author} {\bibfnamefont {A.}~\bibnamefont
  {Santos}},\ }\bibfield  {title} {\enquote {\bibinfo {title} {Class of
  consistent fundamental-measure free energies for hard-sphere mixtures},}\
  }\href {\doibase 10.1103/PhysRevE.86.040102} {\bibfield  {journal} {\bibinfo
  {journal} {Phys. Rev. E}\ }\textbf {\bibinfo {volume} {86}},\ \bibinfo
  {pages} {{040}{102}(R)} (\bibinfo {year} {2012}{\natexlab{b}})}\BibitemShut
  {NoStop}%
\bibitem [{\citenamefont {Santos}\ \emph {et~al.}(2014)\citenamefont {Santos},
  \citenamefont {Yuste}, \citenamefont {{L\'opez de Haro}}, \citenamefont
  {Odriozola},\ and\ \citenamefont {Ogarko}}]{SYHOO14}%
  \BibitemOpen
  \bibfield  {author} {\bibinfo {author} {\bibfnamefont {A.}~\bibnamefont
  {Santos}}, \bibinfo {author} {\bibfnamefont {S.~B.}\ \bibnamefont {Yuste}},
  \bibinfo {author} {\bibfnamefont {M.}~\bibnamefont {{L\'opez de Haro}}},
  \bibinfo {author} {\bibfnamefont {G.}~\bibnamefont {Odriozola}}, \ and\
  \bibinfo {author} {\bibfnamefont {V.}~\bibnamefont {Ogarko}},\ }\bibfield
  {title} {\enquote {\bibinfo {title} {Simple effective rule to estimate the
  jamming packing fraction of polydisperse hard spheres},}\ }\href {\doibase
  10.1103/PhysRevE.89.040302} {\bibfield  {journal} {\bibinfo  {journal} {Phys.
  Rev. E}\ }\textbf {\bibinfo {volume} {89}},\ \bibinfo {pages} {{040}{302}(R)}
  (\bibinfo {year} {2014})}\BibitemShut {NoStop}%
\bibitem [{\citenamefont {Carnahan}\ and\ \citenamefont
  {Starling}(1969)}]{CS69}%
  \BibitemOpen
  \bibfield  {author} {\bibinfo {author} {\bibfnamefont {N.~F.}\ \bibnamefont
  {Carnahan}}\ and\ \bibinfo {author} {\bibfnamefont {K.~E.}\ \bibnamefont
  {Starling}},\ }\bibfield  {title} {\enquote {\bibinfo {title} {Equation of
  state for nonattracting rigid spheres},}\ }\href {\doibase 10.1063/1.1672048}
  {\bibfield  {journal} {\bibinfo  {journal} {J. Chem. Phys.}\ }\textbf
  {\bibinfo {volume} {51}},\ \bibinfo {pages} {635--636} (\bibinfo {year}
  {1969})}\BibitemShut {NoStop}%
\bibitem [{\citenamefont {Boubl{\'i}k}\ and\ \citenamefont
  {Nezbeda}(1986)}]{BN86}%
  \BibitemOpen
  \bibfield  {author} {\bibinfo {author} {\bibfnamefont {T.}~\bibnamefont
  {Boubl{\'i}k}}\ and\ \bibinfo {author} {\bibfnamefont {I.}~\bibnamefont
  {Nezbeda}},\ }\bibfield  {title} {\enquote {\bibinfo {title} {{P-V-T}
  behaviour of hard body fluids. theory and experiment},}\ }\href {\doibase
  10.1135/cccc19862301} {\bibfield  {journal} {\bibinfo  {journal} {Collect.
  Czech. Chem. Commun.}\ }\textbf {\bibinfo {volume} {51}},\ \bibinfo {pages}
  {2301--2432} (\bibinfo {year} {1986})}\BibitemShut {NoStop}%
\bibitem [{\citenamefont {Boubl{\'i}k}(1986)}]{B86}%
  \BibitemOpen
  \bibfield  {author} {\bibinfo {author} {\bibfnamefont {T.}~\bibnamefont
  {Boubl{\'i}k}},\ }\bibfield  {title} {\enquote {\bibinfo {title} {Equations
  of state of hard body fluids},}\ }\href {\doibase 10.1080/00268978600102131}
  {\bibfield  {journal} {\bibinfo  {journal} {Mol. Phys.}\ }\textbf {\bibinfo
  {volume} {59}},\ \bibinfo {pages} {371--380} (\bibinfo {year}
  {1986})}\BibitemShut {NoStop}%
\bibitem [{\citenamefont {Reiss}\ \emph {et~al.}(1960)\citenamefont {Reiss},
  \citenamefont {Frisch}, \citenamefont {Helfand},\ and\ \citenamefont
  {Lebowitz}}]{RFHL60}%
  \BibitemOpen
  \bibfield  {author} {\bibinfo {author} {\bibfnamefont {H.}~\bibnamefont
  {Reiss}}, \bibinfo {author} {\bibfnamefont {H.~L.}\ \bibnamefont {Frisch}},
  \bibinfo {author} {\bibfnamefont {E.}~\bibnamefont {Helfand}}, \ and\
  \bibinfo {author} {\bibfnamefont {J.~L.}\ \bibnamefont {Lebowitz}},\
  }\bibfield  {title} {\enquote {\bibinfo {title} {Aspects of the statistical
  thermodynamics of real fluids},}\ }\href {\doibase 10.1063/1.1700883}
  {\bibfield  {journal} {\bibinfo  {journal} {J. Chem. Phys.}\ }\textbf
  {\bibinfo {volume} {32}},\ \bibinfo {pages} {119--124} (\bibinfo {year}
  {1960})}\BibitemShut {NoStop}%
\bibitem [{\citenamefont {Roth}\ \emph {et~al.}(2002)\citenamefont {Roth},
  \citenamefont {Evans}, \citenamefont {Lang},\ and\ \citenamefont
  {Kahl}}]{RELK02}%
  \BibitemOpen
  \bibfield  {author} {\bibinfo {author} {\bibfnamefont {R.}~\bibnamefont
  {Roth}}, \bibinfo {author} {\bibfnamefont {R.}~\bibnamefont {Evans}},
  \bibinfo {author} {\bibfnamefont {A.}~\bibnamefont {Lang}}, \ and\ \bibinfo
  {author} {\bibfnamefont {G.}~\bibnamefont {Kahl}},\ }\bibfield  {title}
  {\enquote {\bibinfo {title} {Fundamental measure theory for hard-sphere
  mixtures revisited: the {White Bear} version},}\ }\href {\doibase
  10.1088/0953-8984/14/46/313} {\bibfield  {journal} {\bibinfo  {journal} {J.
  Phys.: Condens. Matter}\ }\textbf {\bibinfo {volume} {14}},\ \bibinfo {pages}
  {{12}{063}--{12}{078}} (\bibinfo {year} {2002})}\BibitemShut {NoStop}%
\bibitem [{\citenamefont {Alder}(1964)}]{A64}%
  \BibitemOpen
  \bibfield  {author} {\bibinfo {author} {\bibfnamefont {B.~J.}\ \bibnamefont
  {Alder}},\ }\bibfield  {title} {\enquote {\bibinfo {title} {Studies in
  molecular dynamics. {III. A} mixture of hard spheres},}\ }\href {\doibase
  10.1063/1.1725587} {\bibfield  {journal} {\bibinfo  {journal} {J. Chem.
  Phys.}\ }\textbf {\bibinfo {volume} {40}},\ \bibinfo {pages} {2724--2730}
  (\bibinfo {year} {1964})}\BibitemShut {NoStop}%
\bibitem [{\citenamefont {Rotenberg}(1965)}]{R65}%
  \BibitemOpen
  \bibfield  {author} {\bibinfo {author} {\bibfnamefont {A.}~\bibnamefont
  {Rotenberg}},\ }\bibfield  {title} {\enquote {\bibinfo {title} {Monte {Carlo}
  equation of state for hard spheres in an attractive square well},}\ }\href
  {\doibase 10.1063/1.1696904} {\bibfield  {journal} {\bibinfo  {journal} {J.
  Chem. Phys.}\ }\textbf {\bibinfo {volume} {43}},\ \bibinfo {pages}
  {1198--1201} (\bibinfo {year} {1965})}\BibitemShut {NoStop}%
\bibitem [{\citenamefont {Bannermana}\ and\ \citenamefont {Lue}(2009)}]{BL09}%
  \BibitemOpen
  \bibfield  {author} {\bibinfo {author} {\bibfnamefont {M.~N.}\ \bibnamefont
  {Bannermana}}\ and\ \bibinfo {author} {\bibfnamefont {L.}~\bibnamefont
  {Lue}},\ }\bibfield  {title} {\enquote {\bibinfo {title} {Transport
  properties of highly asymmetric hard-sphere mixtures},}\ }\href {\doibase
  10.1063/1.3120488} {\bibfield  {journal} {\bibinfo  {journal} {J. Chem.
  Phys.}\ }\textbf {\bibinfo {volume} {130}},\ \bibinfo {pages} {164507}
  (\bibinfo {year} {2009})}\BibitemShut {NoStop}%
\bibitem [{\citenamefont {Biben}\ and\ \citenamefont {Hansen}(1991)}]{BH91}%
  \BibitemOpen
  \bibfield  {author} {\bibinfo {author} {\bibfnamefont {T.}~\bibnamefont
  {Biben}}\ and\ \bibinfo {author} {\bibfnamefont {J.-P.}\ \bibnamefont
  {Hansen}},\ }\bibfield  {title} {\enquote {\bibinfo {title} {{Phase
  separation of asymmetric binary hard-sphere fluids}},}\ }\href {\doibase
  10.1103/PhysRevLett.66.2215} {\bibfield  {journal} {\bibinfo  {journal}
  {Phys. Rev. Lett.}\ }\textbf {\bibinfo {volume} {66}},\ \bibinfo {pages}
  {2215--2218} (\bibinfo {year} {1991})}\BibitemShut {NoStop}%
\bibitem [{\citenamefont {Dijkstra}, \citenamefont {van Roij},\ and\
  \citenamefont {Evans}(1999)}]{DRE99b}%
  \BibitemOpen
  \bibfield  {author} {\bibinfo {author} {\bibfnamefont {M.}~\bibnamefont
  {Dijkstra}}, \bibinfo {author} {\bibfnamefont {R.}~\bibnamefont {van Roij}},
  \ and\ \bibinfo {author} {\bibfnamefont {R.}~\bibnamefont {Evans}},\
  }\bibfield  {title} {\enquote {\bibinfo {title} {{Phase diagram of highly
  asymmetric binary hard-sphere mixtures}},}\ }\href {\doibase
  10.1103/PhysRevE.59.5744} {\bibfield  {journal} {\bibinfo  {journal} {Phys.
  Rev. E}\ }\textbf {\bibinfo {volume} {59}},\ \bibinfo {pages} {5744--5771}
  (\bibinfo {year} {1999})}\BibitemShut {NoStop}%
\bibitem [{\citenamefont {Kofke}\ and\ \citenamefont {Bolhuis}(1999)}]{KB99}%
  \BibitemOpen
  \bibfield  {author} {\bibinfo {author} {\bibfnamefont {D.~A.}\ \bibnamefont
  {Kofke}}\ and\ \bibinfo {author} {\bibfnamefont {P.~G.}\ \bibnamefont
  {Bolhuis}},\ }\bibfield  {title} {\enquote {\bibinfo {title} {Freezing of
  polydisperse hard spheres},}\ }\href {\doibase 10.1103/PhysRevE.59.618}
  {\bibfield  {journal} {\bibinfo  {journal} {Phys. Rev. E}\ }\textbf {\bibinfo
  {volume} {59}},\ \bibinfo {pages} {618--622} (\bibinfo {year}
  {1999})}\BibitemShut {NoStop}%
\end{thebibliography}%

\newpage

\appendix
\begin{widetext}
\section*{Supplementary material to the paper ``Chemical potential of a test hard sphere of variable size and free energy in hard-sphere fluid mixtures''}
The following tables give our MD numerical values of $\cc_2$, $\cc_3$, $\beta\mu_1^\ex$, $\beta\mu_2^\ex$, $Z$, and $\beta a^\ex$ for the sets described in Table II of the main text.

\setcounter{table}{0}

\begin{table*}[h]
\caption{Set B1 ($\eta=0.3$, $x_1=0.5$). The number of particles is $N=2048$.}
\begin{ruledtabular}
\begin{tabular}{ddddddd}
\multicolumn{1}{c}{$\quad \sigma_2/\sigma_1$}  & \multicolumn{1}{c}{$\qquad \cc_2$} & \multicolumn{1}{c}{$\qquad\cc_3$} & \multicolumn{1}{c}{$\qquad \beta\mu_1^\ex$} &\multicolumn{1}{c}{$\qquad \beta\mu_2^\ex$}
& \multicolumn{1}{c}{$\qquad Z$} &\multicolumn{1}{c}{$\qquad \beta a^\ex$}\\
\hline
   1.00&2.06908&1.17538& 4.8868& 4.8868& 3.9835& 1.9034\\
   0.95&2.03294&1.19813& 5.1048& 4.6461& 3.9797& 1.8958\\
   0.92&2.11256&1.12554& 5.2558& 4.5154& 3.9734& 1.9122\\
   0.90&1.98199&1.21580& 5.3284& 4.3938& 3.9690& 1.8797\\
   0.85&1.99670&1.16964& 5.5402& 4.1165& 3.9465& 1.8819\\
   0.80&2.00976&1.09738& 5.7326& 3.8234& 3.9162& 1.8681\\
   0.75&1.89714&1.11204& 5.9074& 3.5104& 3.8727& 1.8363\\
   0.70&1.84842&1.05327& 6.0561& 3.2010& 3.8216& 1.8052\\
   0.65&1.82396&0.96196& 6.1845& 2.8920& 3.7567& 1.7816\\
   0.60&1.69779&0.92704& 6.2581& 2.5695& 3.6852& 1.7266\\
   0.58&1.66598&0.89673&6.2826&2.4434&3.6526&1.7104\\
   0.56&1.63709&0.86326&6.3032&2.3245&3.6196&1.6942\\
   0.55&1.62159&0.84608& 6.3099& 2.2655& 3.6029& 1.6848\\
   0.54&1.57884&0.84622&6.3079&2.1998&3.5858&1.6680\\
   0.52&1.54699&0.81336&6.3196&2.0860&3.5507&1.6521\\
   0.50&1.48940&0.79709& 6.3188& 1.9686& 3.5157& 1.6301\\
   0.48&1.46031&0.75675&6.3153&1.8613&3.4782&1.6101\\
   0.46&1.39174&0.74535&6.3039&1.7488&3.4411&1.5852\\
   0.45&1.33356&0.75359& 6.2882& 1.6883& 3.4222& 1.5660\\
   0.44&1.36578&0.70187&6.2859&1.6491&3.4036&1.5639\\
   0.42&1.32107&0.67216&6.2635&1.5495&3.3652&1.5413\\
   0.40&1.26705&0.64878& 6.2363& 1.4553& 3.3277& 1.5162\\
   0.35&1.13331&0.58986& 6.1459& 1.2280& 3.2313& 1.4556\\
   0.30&1.03243&0.51564& 6.0461& 1.0413& 3.1392& 1.4004\\
   0.25&0.98581&0.41657& 5.9317& 0.8773& 3.0483& 1.3562\\
   0.20&0.84588&0.38120& 5.7956& 0.7310& 2.9655& 1.2984\\
   0.15&0.76751&0.31920& 5.6642& 0.6108& 2.8881& 1.2511\\
   0.10&0.66038&0.28290& 5.5335& 0.5103& 2.8189& 1.2046\\
   0.05&0.57719&0.24216& 5.4172& 0.4267& 2.7583& 1.1615\\
   0.03&0.56247&0.22079& 5.3807& 0.3972& 2.7353& 1.1537\\
\end{tabular}
\end{ruledtabular}
\end{table*}

\begin{table*}
\caption{Set B2 ($\eta=0.3$, $\sigma_2/\sigma_1=0.5$). The number of particles is $N=2048$.}
\begin{ruledtabular}
\begin{tabular}{ddddddd}
\multicolumn{1}{c}{$\quad x_1$}  & \multicolumn{1}{c}{$\qquad \cc_2$} & \multicolumn{1}{c}{$\qquad\cc_3$} & \multicolumn{1}{c}{$\qquad \beta\mu_1^\ex$} &\multicolumn{1}{c}{$\qquad \beta\mu_2^\ex$}
& \multicolumn{1}{c}{$\qquad Z$} &\multicolumn{1}{c}{$\qquad \beta a^\ex$}\\
\hline
    0.9501953&1.98839&1.13722& 4.9686& 1.6799&3.9317& 1.8731\\
    0.8500977&1.86781&1.03654& 5.1623& 1.7228&3.8300& 1.8168\\
    0.7500000&1.70284&0.98433& 5.3873& 1.7649&3.7316& 1.7501\\
    0.6499023&1.64226&0.87195& 5.6898& 1.8375&3.6391& 1.7020\\
    0.6000977&1.60244&0.83390&5.8732&1.8784&3.5957&1.6800\\
    0.5800781&1.60617&0.80216&5.9491&1.8995&3.5785&1.6702\\
    0.5600586&1.51196&0.85046&6.0342&1.9022&3.5621&1.6543\\
    0.5498047&1.55590&0.80386& 6.0805& 1.9230&3.5536& 1.6551\\
    0.5400391&1.50876&0.82624&6.1204&1.9237&3.5456&1.6445\\
    0.5200195&1.53331&0.78505&6.2182&1.9532&3.5299&1.6412\\
    0.5000000&1.48851&0.79647&6.3194&1.9685&3.5149&1.6291\\
    0.4799805&1.48564&0.77792&6.4286&1.9945&3.5002&1.6226\\
    0.4599609&1.46742&0.77158&6.5443&2.0182&3.4865&1.6135\\
    0.4501953&1.41707&0.79751& 6.5998& 2.0200&3.4797& 1.6021\\
    0.4399414&1.46557&0.75398&6.66652&2.0472&3.4733&1.6061\\
    0.4199219&1.42040&0.77082&6.8037&2.0679&3.4605&1.5961\\
    0.3999023&	1.43297&0.74574&6.9449&2.1041&3.4485&1.5914\\
    0.3500977&1.40077&0.73858& 7.3595& 2.1892&3.4239& 1.5755\\
    0.2500000&1.41077&0.71527& 8.5344& 2.4440&3.4012& 1.5654\\
    0.2299805&1.43865&0.70266& 8.8465& 2.5178&3.4030& 1.5703\\
    0.1899414&1.41413&0.74708& 9.6314& 2.6652&3.4160& 1.5724\\
    0.1601562&1.44192&0.76495&10.3556& 2.8152&3.4365& 1.5864\\
    0.1298828&1.51050&0.77285&11.2475& 3.0115&3.4706& 1.6106\\
    0.0698242&1.63740&0.88691&13.9630& 3.5562&3.6007& 1.6821\\
    0.0498047&1.70936&0.94941&15.3171& 3.8241&3.6737& 1.7228\\
    0.0297852&1.87025&0.98533&16.9473& 4.1815&3.7709& 1.7909\\
\end{tabular}
\end{ruledtabular}
\end{table*}

\begin{table*}
\caption{Set B3 ($\eta=0.3$, $\sigma_2/\sigma_1=0.3$). The number of particles is $N=2048$ and $N=4000$ (values with a dagger).}
\begin{ruledtabular}
\begin{tabular}{ddddddd}
\multicolumn{1}{c}{$\quad x_1$}  & \multicolumn{1}{c}{$\qquad \cc_2$} & \multicolumn{1}{c}{$\qquad\cc_3$} & \multicolumn{1}{c}{$\qquad \beta\mu_1^\ex$} &\multicolumn{1}{c}{$\qquad \beta\mu_2^\ex$}
& \multicolumn{1}{c}{$\qquad Z$} &\multicolumn{1}{c}{$\qquad \beta a^\ex$}\\
\hline
    0.9501953&1.94452&1.09235& 4.9492& 0.9643&3.8973& 1.8535\\
    0.9000000^\dagger&1.80039^\dagger&1.02976^\dagger& 5.0132^\dagger& 0.9670^\dagger&3.8113^\dagger& 1.7973^\dagger\\
    0.8500977&1.73159&0.91791& 5.0981& 0.9757&3.7251& 1.7550\\
    0.8000000^\dagger&1.55882^\dagger&0.88967^\dagger& 5.1689^\dagger& 0.9759^\dagger&3.6396^\dagger& 1.6907^\dagger\\
    0.7500000&1.52024&0.77348& 5.2802& 0.9886&3.5543& 1.6531\\
    0.7000000^\dagger&1.39952^\dagger&0.72407^\dagger& 5.3878^\dagger& 0.9942^\dagger&3.4696^\dagger& 1.6001^\dagger\\
    0.6500000^\dagger&1.26365^\dagger&0.69151^\dagger& 5.5090^\dagger& 0.9982^\dagger&3.3857^\dagger& 1.5445^\dagger\\
    0.6499023&1.26182&0.69116& 5.5058& 0.9979&3.3851& 1.5425\\
    0.5498047&1.16797&0.52882& 5.8437& 1.0304&3.2190& 1.4577\\
    0.5000000^\dagger&1.08306^\dagger&0.48389^\dagger& 6.0467^\dagger& 1.0443^\dagger&3.1384^\dagger& 1.4071^\dagger\\
    0.4501953&1.00564&0.44403& 6.3035& 1.0618&3.0584& 1.3632\\
    0.4000000^\dagger&0.90943^\dagger&0.41939^\dagger& 6.6120^\dagger& 1.0788^\dagger&2.9805^\dagger& 1.3116^\dagger\\
    0.3500977&0.88918&0.36094& 7.0073& 1.1149&2.9056& 1.2722\\
    0.3000000^\dagger&0.80847^\dagger&0.34576^\dagger& 7.5351^\dagger& 1.1459^\dagger&2.8353^\dagger& 1.2273^\dagger\\
    0.2500000&0.73134&0.33962& 8.2775& 1.1871&2.7707& 1.1890\\
    0.2500000^\dagger&0.74275^\dagger&0.33413^\dagger& 8.2768^\dagger& 1.1903^\dagger&2.7710^\dagger& 1.1910^\dagger\\
    0.2299805&0.73275&0.32354& 8.6418& 1.2165&2.7476& 1.1766\\
    0.2100000^\dagger&0.73932^\dagger&0.30735^\dagger& 9.0604^\dagger& 1.2513^\dagger&2.7274^\dagger& 1.1638^\dagger\\
    0.1900000^\dagger&0.69508^\dagger& 0.32240^\dagger& 9.6306^\dagger&1.2762^\dagger&2.7090^\dagger& 1.1545^\dagger\\
    0.1899414&0.72012&0.30935& 9.6054& 1.2840&2.7090& 1.1556\\
    0.1601562&0.72374&0.30196&10.5903& 1.3540&2.6889& 1.1444\\
    0.1298828&0.70967&0.31826&12.0734& 1.4420&2.6820& 1.1409\\
    0.1000000^\dagger&0.73302^\dagger&0.33858^\dagger&14.2677^\dagger& 1.5807^\dagger&2.6977^\dagger& 1.1517^\dagger\\
    0.0698242&0.81544&0.37403&17.9438& 1.8213&2.7575& 1.1895\\
    0.0498047&0.91456&0.42801&22.2170& 2.0902&2.8486& 1.2440\\
    0.0300000^\dagger&1.14103^\dagger&0.51230^\dagger&29.6029^\dagger& 2.5767^\dagger&3.0319^\dagger& 1.3556^\dagger\\
    0.0297852&1.10773&0.53651&30.0497& 2.5705&3.0345& 1.3545\\
    0.0100000^\dagger&1.57822^\dagger&0.79039^\dagger&47.8872^\dagger& 3.6420^\dagger&3.4728^\dagger& 1.6116^\dagger\\
\end{tabular}
\end{ruledtabular}
\end{table*}

\begin{table*}
\caption{Set B4 ($x_1=0.5$, $\sigma_2/\sigma_1=0.5$). The number of particles is $N=2048$ and $N=4000$ (values with a dagger).}
\begin{ruledtabular}
\begin{tabular}{ddddddd}
\multicolumn{1}{c}{$\quad \eta$}  & \multicolumn{1}{c}{$\qquad \cc_2$} & \multicolumn{1}{c}{$\qquad\cc_3$} & \multicolumn{1}{c}{$\qquad \beta\mu_1^\ex$} &\multicolumn{1}{c}{$\qquad \beta\mu_2^\ex$}
& \multicolumn{1}{c}{$\qquad Z$} &\multicolumn{1}{c}{$\qquad \beta a^\ex$}\\
\hline
   0.100&0.28617&0.10958& 1.2442& 0.4502&1.4505&0.3967\\
   0.100^{\dagger}&0.28743^\dagger&0.109003^\dagger&1.2451^\dagger&0.4506^\dagger&1.4507^\dagger&0.3972^\dagger\\
   0.150&0.49061&0.20056& 2.0983& 0.7341&1.7769&0.6393\\
   0.150^{\dagger}&0.50223^\dagger&0.19551^\dagger&2.1070^\dagger&0.7378^\dagger&1.7773^\dagger&0.6451^\dagger\\
   0.200&0.75177&0.32939& 3.1737& 1.0715&2.2025&0.9201\\
   0.200^{\dagger}&0.76881^\dagger&0.31960^\dagger&3.1808^\dagger&1.0762^\dagger&2.2026^\dagger&0.9259^\dagger\\
   0.250&1.08291&0.51532&4.5455& 1.4773&2.7639&1.2475\\
   0.250^{\dagger}&1.05070^\dagger&0.53386^\dagger&4.5322^\dagger&1.4684^\dagger&2.7642^\dagger&1.2361^\dagger\\
   0.260&1.15943&0.56039& 4.8618& 1.5680&2.8966&1.3183\\
   0.270&1.23403&0.61354& 5.1957& 1.6614&3.0378&1.3908\\
   0.280&1.31562&0.66966& 5.5510& 1.7598&3.1873&1.4681\\
   0.290&1.40763&0.72396& 5.9225& 1.8634&3.3461&1.5468\\
   0.300&1.48805&0.79620& 6.3180& 1.9682&3.5148&1.6284\\
   0.300^{\dagger}&1.59864^\dagger&0.71777^\dagger&6.3287^\dagger&1.99417^\dagger&3.5155^\dagger&1.6459^\dagger\\
   0.310&1.61584&0.84022& 6.7329& 2.0870&3.6943&1.7156\\
   0.320&1.71326&0.91762& 7.1752& 2.2033&3.8854&1.8038\\
   0.330&1.81236&1.00418& 7.6445& 2.3244&4.0885&1.8960\\
   0.340&1.92517&1.09210& 8.1439& 2.4533&4.3049&1.9937\\
   0.350&2.06434&1.17007& 8.6696& 2.5924&4.5360&2.0948\\
   0.350^{\dagger}&2.09213^\dagger&1.14644^\dagger&8.6625^\dagger&2.5977^\dagger&4.5362^\dagger&2.0939^\dagger\\
   0.375&2.37893&1.44563&10.1259& 2.9556&5.1840&2.3568\\
   0.375^{\dagger}&2.46590^\dagger&1.36653^\dagger&10.09303^\dagger&2.9709^\dagger&5.1841^\dagger&2.3479^\dagger\\
   0.400&2.81908&1.71536&11.8108& 3.3831&5.9516&2.6454\\
   0.400^{\dagger}&2.94999^\dagger&1.59050^\dagger&11.7475^\dagger&3.4043^\dagger&5.9517^\dagger&2.6243^\dagger\\
   0.425&3.28682&2.08366&13.7994& 3.8635&6.8647&2.9667\\
   0.425^{\dagger}&3.21021^\dagger&2.15112^\dagger&13.82312^\dagger&3.8493^\dagger&6.8646^\dagger&2.9717^\dagger\\
   0.450&3.73708&2.62871&16.1998& 4.4013&7.9608&3.3397\\
   0.450^{\dagger}&3.80929^\dagger&2.54949^\dagger&16.1404^\dagger&4.4099^\dagger&7.9623^\dagger&3.3129^\dagger\\
   0.475&4.31657&3.26421&19.0715& 5.0379&9.2853&3.7695\\
   0.475^{\dagger}&4.52378^\dagger&3.01780^\dagger&18.8558^\dagger&5.0570^\dagger&9.2844^\dagger&3.6720^\dagger\\
   0.500&4.78244&4.33293&22.7993& 5.7692&10.9004&4.3838\\
   0.500^{\dagger}&4.84377^\dagger&4.264407^\dagger&22.74587^\dagger&5.77617^\dagger&10.9029^\dagger&4.3581^\dagger\\
\end{tabular}
\end{ruledtabular}
\end{table*}

\begin{table*}
\caption{Set B5 ($x_1=0.5$, $\sigma_2/\sigma_1=2-\sqrt{3}$). The number of particles is $N=2048$ and $N=4000$ (values with a dagger).}
\begin{ruledtabular}
\begin{tabular}{ddddddd}
\multicolumn{1}{c}{$\quad \eta$}  & \multicolumn{1}{c}{$\qquad \cc_2$} & \multicolumn{1}{c}{$\qquad\cc_3$} & \multicolumn{1}{c}{$\qquad \beta\mu_1^\ex$} &\multicolumn{1}{c}{$\qquad \beta\mu_2^\ex$}
& \multicolumn{1}{c}{$\qquad Z$} &\multicolumn{1}{c}{$\qquad \beta a^\ex$}\\
\hline
   0.050&0.08427&0.02928& 0.5419& 0.1130&1.1685& 0.1590\\
   0.100&0.19035&0.06931& 1.2015& 0.2385&1.3817& 0.3384\\
   0.150&0.31716&0.12700& 2.0067& 0.3779&1.6542& 0.5382\\
   0.200&0.48130&0.20632& 3.0190& 0.5360&2.0063& 0.7712\\
   0.250&0.69476&0.31592& 4.3077& 0.7174&2.4674& 1.0451\\
   0.300&1.02791&0.43704& 5.9813& 0.9356&3.0797& 1.3787\\
   0.325&1.15013&0.56222& 6.9799& 1.0479&3.4616& 1.5524\\
   0.350&1.31316&0.69987& 8.1433& 1.1734&3.9070& 1.7514\\
   0.375&1.58764&0.80276& 9.4634& 1.3214&4.4296& 1.9628\\
   0.400&1.77357&1.02879&11.0641& 1.4688&5.0463& 2.2202\\
   0.425&2.12074&1.19419&12.8482& 1.6472&5.7792& 2.4685\\
   0.450&2.42204&1.48873&15.0476& 1.8345&6.6531& 2.7879\\
   0.475&2.78091&1.83842&17.6325& 2.0447&7.7086& 3.1300\\
   0.475^{\dagger}&2.91674^\dagger&1.70959^\dagger&17.4648^\dagger&2.0592^\dagger&7.7090^\dagger&3.0531^\dagger\\
   0.500&3.47964&1.99169&20.3217& 2.3104&8.9883& 3.3277\\
   0.500^{\dagger}&3.22734^\dagger&2.27104^\dagger&20.7902^\dagger&2.2864^\dagger&8.9914^\dagger&3.5469^\dagger\\
   \end{tabular}
\end{ruledtabular}
\end{table*}

\begin{table*}
\caption{Set B6 [$\eta=(x_1+0.1)/2$, $\sigma_2/\sigma_1=x_1$). The number of particles is $N=2048$ and $N=4000$ (values with a dagger).}
\begin{ruledtabular}
\begin{tabular}{ddddddd}
\multicolumn{1}{c}{$\quad x_1$}  & \multicolumn{1}{c}{$\qquad \cc_2$} & \multicolumn{1}{c}{$\qquad\cc_3$} & \multicolumn{1}{c}{$\qquad \beta\mu_1^\ex$} &\multicolumn{1}{c}{$\qquad \beta\mu_2^\ex$}
& \multicolumn{1}{c}{$\qquad Z$} &\multicolumn{1}{c}{$\qquad \beta a^\ex$}\\
\hline
    0.2000000^{\dagger}&0.14556^\dagger&0.04869^\dagger&2.9243^\dagger&0.3348^\dagger&1.4595^\dagger&0.3933^\dagger\\
    0.2001953&0.13840&0.05052& 2.9060& 0.3329&1.4594& 0.3885\\
    0.2500000&0.24703&0.09429& 3.3307& 0.4711&1.6546& 0.5314\\
    0.2500000^{\dagger}&0.23942^\dagger&0.096682^\dagger&3.31962^\dagger&0.4691^\dagger&1.6546^\dagger&0.5271^\dagger\\
    0.2998047&0.39853&0.15965& 3.8145& 0.6498&1.8987& 0.6999\\
    0.3000000^{\dagger}&0.40096^\dagger&0.15899^\dagger&3.8170^\dagger&0.6504^\dagger&1.8990^\dagger&0.7014^\dagger\\
    0.3500000^{\dagger}&0.58094^\dagger&0.26047^\dagger&4.3394^\dagger&0.8729^\dagger&2.1989^\dagger&0.8873^\dagger\\
    0.3500977&0.60107&0.25127& 4.3512& 0.8781&2.1988& 0.8952\\
    0.3999020&0.84887&0.38120& 4.9465& 1.1650&2.5619& 1.1153\\
    0.4199220&1.09211&0.40359& 5.2075& 1.2798&2.7268& 1.2023\\
    0.4399410&1.12566&0.49191& 5.4732& 1.4559&2.9040& 1.3192\\
    0.4599610&1.19642&0.60992& 5.7357& 1.6013&3.0941& 1.4088\\
    0.4799800&1.40050&0.65655& 6.0315& 1.7920&3.2970& 1.5298\\
    0.5000000&1.48839&0.79729& 6.3212& 1.9687&3.5152& 1.6300\\
    0.5000000^{\dagger}&1.59864^\dagger&0.71777^\dagger&6.3287^\dagger&1.9941^\dagger&3.5155^\dagger&1.6459^\dagger\\
    0.5200200&1.74562&0.83546& 6.6418& 2.2009&3.7478& 1.7625\\
    0.5400390&1.82343&1.02186& 6.9582& 2.4085&3.9966& 1.8689\\
    0.5600590&1.91924&1.21908& 7.2946& 2.6405&4.2623& 1.9848\\
    0.5800780&2.13553&1.34415& 7.6532& 2.9206&4.5463& 2.1196\\
    0.6000000^{\dagger}&2.39702^\dagger&1.44975^\dagger&8.0249^\dagger&3.2327^\dagger&4.8502^\dagger&2.2578^\dagger\\
    0.6000977&2.36457&1.47852& 8.0267& 3.2264&4.8491& 2.2580\\
    0.6499023&2.90866&1.94093& 9.0455& 4.1063&5.6993& 2.6170\\
    0.6500000^{\dagger}&2.97090^\dagger&1.87456^\dagger&9.0275^\dagger&4.1132^\dagger&5.7005^\dagger&2.6069^\dagger\\
    0.7000000^{\dagger}&3.74195^\dagger&2.28598^\dagger&10.1729^\dagger&5.2424^\dagger&6.7018^\dagger&2.9920^\dagger\\
    0.7001953&3.56817&2.47073&10.2070& 5.2231&6.7014& 3.0113\\
    0.7500000&4.27246&3.15574&11.5529& 6.6349&7.8859& 3.4375\\
    0.7500000^{\dagger}&4.35285^\dagger&3.056507^\dagger&11.5239^\dagger&6.6355^\dagger&7.8858^\dagger&3.4161^\dagger\\
    0.7998047&4.91892&4.18517&13.1909& 8.4557&9.2931& 3.9499\\
    0.800000^{\dagger}&4.82801^\dagger&4.30670^\dagger&13.2285^\dagger&8.4623^\dagger&9.2915^\dagger&3.9838^\dagger\\
    0.8500000^{\dagger}&6.22450^\dagger&4.67071^\dagger&14.9211^\dagger&10.7694^\dagger&10.9688^\dagger&4.3295^\dagger\\
    0.8500977&5.90649&5.11906&15.0673&10.8230&10.9699& 4.4611\\
    0.8750000&6.63409&5.38660&16.0513&12.2215&11.9316& 4.6410\\
    0.8750000^{\dagger}&6.54644^\dagger&5.50991^\dagger&16.0901^\dagger&12.2388^\dagger&11.9308^\dagger&4.6779^\dagger\\
    \end{tabular}
\end{ruledtabular}
\end{table*}

\begin{table*}
\caption{Set TH1 ($\eta=0.3$). The number of particles is $N=2048$.}
\begin{ruledtabular}
\begin{tabular}{ddddddd}
\multicolumn{1}{c}{$\quad \sigma_{\min}/\sigma_{\max}$}  & \multicolumn{1}{c}{$\qquad \cc_2$} & \multicolumn{1}{c}{$\qquad\cc_3$} & \multicolumn{1}{c}{$\qquad \beta\mu^\ex(\sigma_{\max})$} &\multicolumn{1}{c}{$\qquad \beta\mu^\ex(\sigma_{\min})$}
& \multicolumn{1}{c}{$\qquad Z$} &\multicolumn{1}{c}{$\qquad \beta a^\ex$}\\
\hline
   0.1&1.20124&0.57098& 9.4646& 0.5710&3.2475& 1.4737\\
   0.3&1.54256&0.83242& 8.7158& 1.2713&3.5611& 1.6540\\
   0.4&1.71478&0.92390& 8.1931& 1.7472&3.6890& 1.7338\\
   0.5&1.86766&0.98753& 7.6153& 2.2788&3.7933& 1.7975\\
   0.6&1.92881&1.07924& 7.0208& 2.8229&3.8709& 1.8368\\
   0.7&1.96624&1.15255& 6.4359& 3.3709&3.9253& 1.8663\\
   0.8&1.97014&1.21168& 5.8674& 3.8975&3.9598& 1.8770\\
   0.9&2.06757&1.17078& 5.3638& 4.4234&3.9780& 1.9055\\
   1.0&2.01855&1.21857& 4.8795& 4.8795&3.9833& 1.8962\\
\end{tabular}
\end{ruledtabular}
\end{table*}

\begin{table*}
\caption{Set TH2 ($\sigma_{\min}/\sigma_{\max}=0$). The number of particles is $N=2048$ and $N=4000$ (values with a dagger).}
\begin{ruledtabular}
\begin{tabular}{dddddd}
\multicolumn{1}{c}{$\quad \eta$}  & \multicolumn{1}{c}{$\qquad \cc_2$} & \multicolumn{1}{c}{$\qquad\cc_3$} & \multicolumn{1}{c}{$\qquad \beta\mu^\ex(\sigma_{\max})$} & \multicolumn{1}{c}{$\qquad Z$} &\multicolumn{1}{c}{$\qquad \beta a^\ex$}\\
\hline
   0.050&0.08419&0.02933& 0.8325& 1.1685& 0.1590\\
   0.100&0.18967&0.06964& 1.8640& 1.3818& 0.3381\\
   0.150&0.32996&0.12260& 3.1662& 1.6544& 0.5463\\
   0.200&0.48228&0.20650& 4.7998& 2.0067& 0.7726\\
   0.250&0.68968&0.31849& 6.9209& 2.4681& 1.0430\\
   0.300&0.98093&0.46445& 9.7006& 3.0808& 1.3700\\
   0.325&1.11544&0.58353&11.4373& 3.4625& 1.5481\\
   0.350&1.35507&0.67250&13.3712& 3.9080& 1.7517\\
   0.375&1.61024&0.78977&15.6130& 4.4318& 1.9651\\
   0.400&1.83094&0.98660&18.3747& 5.0486& 2.2103\\
   0.425&2.10309&1.21357&21.6077& 5.7815& 2.4818\\
   0.450&2.44410&1.46965&25.3767& 6.6576& 2.7752\\
   0.475&2.81882&1.80811&29.9706& 7.7137& 3.1153\\
   0.475^{\dagger}&2.82256^\dagger&1.79834^\dagger&29.9235^\dagger&7.7116^\dagger&3.1028^\dagger\\
   0.500&3.23142&2.27006&35.7391& 8.9965& 3.5458\\
   0.500^{\dagger}&3.12356^\dagger&2.37451^\dagger&36.1625^\dagger&8.9957^\dagger&3.6117^\dagger\\
\end{tabular}
\end{ruledtabular}
\end{table*}

\end{widetext}

\end{document}